%% file: main.tex
\documentclass[reprint, 10pt,
prx,
aps, 
amsmath,amssymb,
superscriptaddress,
floatfix
]{revtex4-2} 

\usepackage{bm}   
\usepackage{graphicx}   
\usepackage[
  colorlinks=true,
  urlcolor=blue,
  linkcolor=black,
  citecolor=black
]{hyperref}

\usepackage[a4paper, top=2.2cm,bottom=2.5cm,left=1.5cm,right=1.5cm]{geometry}

\usepackage[english]{babel}
\usepackage[utf8]{inputenc}
\usepackage{setspace}
\allowdisplaybreaks

\usepackage{times}
\usepackage{braket}          
\newcommand{\commas}[1]{``{#1}''}
\newenvironment{eqnlist}{\subequations\align}{\endalign\endsubequations}

\begin{document}


\title{Implementation of single-qubit gates via parametric modulation in the Majorana transmon}
\author{E. Lupo}
\affiliation{Advanced Technology Institute and Department of Physics, University of Surrey, Guildford, GU2 7XH, UK}
\author{E. Grosfeld}
\affiliation{Department of Physics, Ben-Gurion University of the Negev, Beer-Sheva 84105, Israel}
\author{E. Ginossar}
\affiliation{Advanced Technology Institute and Department of Physics, University of Surrey, Guildford, GU2 7XH, UK}

\begin{abstract}
\setstretch{1.02}
We present a voltage gate-based method for controlling the Majorana transmon, using a sinusoidal modulation of the induced offset charge $n_g$. Working in the transmon regime and in the instantaneous eigenstates basis, we find the time evolution under this protocol that realises tunable $X$-$Z$ rotations. We optimise the parameters of the system for different single-qubit gates in both the laboratory frame and the qubit rotating frame, obtaining qubit control errors $1-\mathcal{F}$ smaller than $\sim2\times 10^{-4}$. In addition to this, we conduct an analysis of the effects of the charge noise, assuming wide-band $1/f$ additive noise in $n_g$, both for the free and the driven evolutions. For the free evolution, the relaxation and dephasing rates are calculated perturbatively, obtaining long dephasing times of the order of milliseconds at the system's sweet-spots. For the driven case, the average fidelity for the $X$-gate is obtained via a numerical simulation, demonstrating remarkable resilience. \\ \\
\end{abstract}

\maketitle
\setstretch{1.03}

\section{Introduction}

Superconducting circuits based on transmon qubit \cite{Nakamura1999, Blais2004, Koch2007} are among the most studied platforms for quantum information processing, showing promising results in terms of coherent control and scalability \cite{Krantz2019, Kjaergaard2020}. Following their huge success, new designs \cite{Jin2015,Yan2016, Nguyen2019, Grimm2020, Gyenis2021} are continuously being proposed to accelerate the development of either fault-tolerant \cite{Preskill1998} or noisy intermediate-scale quantum (NISQ) computers \cite{Kjaergaard2020}. These include ideas for hybrid platforms where new solid-state elements are embedded in the superconducting circuit to improve both their controllability and their sensitivity to various decoherence mechanism \cite{Larsen2015, Casparis2016, Luthi2018, Casparis2018gatemon2DEG, Casparis2019, schmitt2020, Hays2021}.
Specifically, embedding topological superconductors, which harbour collective topological excitations such as Majorana zero modes (MZMs) \cite{ReadGreen2000, Kitaev2001, Alicea2012}, into superconducting circuit architectures, could lead to new ways for detecting them and potentially to new qubit designs. Either semiconducting nanowires with strong spin-orbit coupling or topological insulator nanowires and nanoribbons represent promising candidates for the creation and manipulation of MZMs \cite{Fu2008, Akhmerov2009interferometryMFinTI, Lutchyn2010, Oreg2010, Cook2011, Huang2017, Manousakis2017, Yavilberg2019, Avila2020models, Avila2020oscill, Prada2020, Tan2021}. Several experimental studies show progress towards building topological Josephson junctions using Indium-based semiconducting materials like InAs or InSb \cite{Mourik2012, Rokhinson2012, Deng2016, Nichele2017, Laroche2019, Vaitieknas2020} or Bismuth-based topological insulators such as Bi$_2$Se$_3$ and Bi$_2$Te$_3$  \cite{Williams2012, Veldhorst2012, Galletti2014, Kurter2015, Charpentier2017}.

Because of their intrinsic topological protection, the introduction of MZMs in hybrid devices may be proven useful for quantum information processing \cite{Alicea2011, Flensberg2011, Hassler2011,VanHeck2012,Ginossar2014,Yavilberg2015,Plugge2017, Karzig2017,Bauer2018, Li2018}. Several works propose qubit designs and protocols for implementing topologically protected braiding operations, either by spatially moving the Majorana modes via electrostatic gates \cite{Alicea2011,Bauer2018} or by applying equivalent methods such as Coulomb-assisted braiding \cite{Hassler2011, VanHeck2012} and projective-measurement based protocols \cite{Flensberg2011, Plugge2017, Karzig2017}. These schemes rely on the precise adiabatic control of electrostatic gates, effective charging energies or tunnel couplings to quantum dots, exposing the stored quantum information to multiple decoherence effects related to the use of nearby electrodes \cite{Scheurer2013, Mishmash2020,Munk2020,Steiner2020,Maman2020}. In addition to this, from an engineering point of view these braiding-based proposed qubits are highly complex devices, hence a simpler qubit design would be beneficial. Following this direction, an alternative approach for the realisation of Majorana-based hybrid qubits \cite{Ginossar2014,Yavilberg2015, Li2018, Vayrynen2021} introduces a weak interaction energy $E_M$ originating from the partial overlap of neighbouring MZMs. When operating in the \commas{Majorana transmon} regime $E_M \ll E_C \ll E_J$, with $E_C$ and $E_J$ the charging and Josephson energies of the circuit, the spectrum resembles that of a harmonic oscillator with each level doubled and each doublet well separated from the others \cite{Ginossar2014,Yavilberg2015}. The introduction of the coupling $E_M$ inevitably sacrifices the full topological protection of the system. However, the electrostatic dipole coupling between levels of the same doublet is exponentially suppressed, making the lowest two energy levels protected from radiative decay. The embedding of this system in a circuit QED architecture has been shown to be a promising tool for the qubit readout \cite{Yavilberg2015, Smith2020} or, in general, for a reliable detection method of the Majorana modes \cite{Yavilberg2019, Avila2020models, Avila2020oscill}.
Recent works \cite{Yavilberg2015, Li2018, Wang2018} also started to explore the challenge of controlling the qubit state. The high anharmonicity in the spectrum in principle means that fast gates could be obtained without introducing unwanted transitions to higher-energy levels; hence, the use of a transmission line resonator \cite{Yavilberg2015} has been proposed for the control of the qubit. Nonetheless, because of the said vanishing intradoublet dipole couplings and the symmetry of the interdoublet ones, a two-tone drive that exploits the transitions to higher doublets needs to be used to obtain coherent oscillations, leading to long gate times \cite{Yavilberg2015}. In other theoretical work \cite{Li2018} on Majorana-based qubits it was shown that coherent oscillations can arise from a combination of different parameters' switches starting and ending in the charging regime $E_C\gg E_J$. It was also recently suggested \cite{Wang2018} to use multiple sweeps through an avoided crossing of the spectrum for the manipulation of MZMs in Josephson junctions, exploiting the Landau-Zener-St\"{u}ckelberg interferometry effect. Other types of time-modulated parametric manipulations have also been considered for qubit readout in works on similar Majorana-based setups \cite{Grimsmo2019,Ohm2015}.
This raises the intriguing possibility that similar methods could be used to resolve the pending challenge of how to implement a universal set of fast, single-qubit gates in the Majorana transmon.

In this work, we propose a precise and systematic voltage gate-based method for the control of the Majorana transmon qubit exploiting the dynamical modulation of the induced offset charge parameter $n_g$, where $n_g$ is measured in units of Cooper pair charge $2e$, and consequentially analyse it under the effect of charge noise. The periodic modulation of such a parameter, in the Majorana transmon regime defined above, introduces a nonlinear driving term in the Hamiltonian of the lowest doublet's subspace that we show can be used to realise a set of high-fidelity single-qubit gates. Furthermore, we show that even if the parameter modulation itself represents a potential decoherence channel of the system, the analysis conducted for both the free and driven evolution under intrinsic $1/f$ charge noise effect yields long relaxation and dephasing times. This shows how the modulation of this specific internal parameter of the system, which is experimentally controlled via an external voltage bias $V_g\propto n_g$, can represent a possible alternative to the microwave control or other gate-based schemes. 

The high anharmonicity of the system allows us to restrict the dynamics to the lowest-energy doublet of the spectrum. We employ the basis of the instantaneous eigenstates of this subspace and, making use of the counter-rotating hybridised rotating-wave method \cite{Lu2012}, we find the solutions of the model beyond the rotating-wave approximation regime and show that a sinusoidal modulation of $n_g(t)$ leads to an effective rotation of the qubit about an axis lying in the $x$-$z$ plane. This rotation axis is indeed tunable via the other internal parameters $E_M$ and $E_J$, thus making any single-qubit gate involving an $X$ or $Z$ rotation possible. The use of the instantaneous eigenstates of the Hamiltonian as the computational basis for the qubit instead of the parity states of the parent superconducting system obviates the need for tuning the interaction energy $E_M$ to the charging regime  \cite{Li2018} for the initialisation and readout stages, reducing any unwanted transitions. When the parameters involved are optimised, we obtain an implementation of an $X$ gate with control error (infidelity $1-\mathcal{F}_G$) $\sim 10^{-4}$, which is in accordance with the threshold for implementing fault-tolerant quantum computation \cite{Preskill1998}. The hybridisation due to $E_M$ makes the system itself sensitive to modulations of $n_g(t)$ over the range $[0,1/2]$; hence, we study the effect of additive, Gaussian $1/f$ noise fluctuations on the offset charge parameter $n_g$. Other sources of charge noise can of course be present in the system and have been analysed elsewhere \cite{Knapp2018, Karzig2021}; however, here we decide to focus on noise which is intrinsic to the system and happening on relevant timescales. We use a perturbative analysis of the Liouville equation to derive the relaxation and dephasing rates for the free evolution and we find that the system presents a sweet-spot at $n_g=0$, with dephasing time $T_\phi \sim 1.4-14\ \text{ms}$, that is typically beyond the current state-of-the-art transmon \cite{Krantz2019}. This is also significant because $n_g=0$ represents the initial and final value for the gate protocol, and thus also the idle point when implementing a quantum algorithm. For the driven evolution, the additive noise is simulated numerically during the protocol, leading to a reduction of the $X$ gate fidelity smaller than $0.01$ percent. The protocol shows also low sensitivity to systematic errors on the initial value of the offset charge $ n_g(0)$ and on the parameters $E_M$ and $E_J$.

 The paper is structured as follows. Section \ref{sec:TheModel} presents an overview of the system and the protocol chosen. In Section \ref{Sec:Dynamics} an effective rotation of the qubit about a tunable axis of the $x$-$z$ plane is derived. Section \ref{sec:OtherGates} extends this method for the implementation of various single-qubit gates. Finally, section \ref{sec:Noise} presents the effect of the $1/f$ charge noise on the system. In section \ref{sec:Conclusions} we summarise and conclude.
 \begin{figure}[t]
\includegraphics[width=\linewidth]{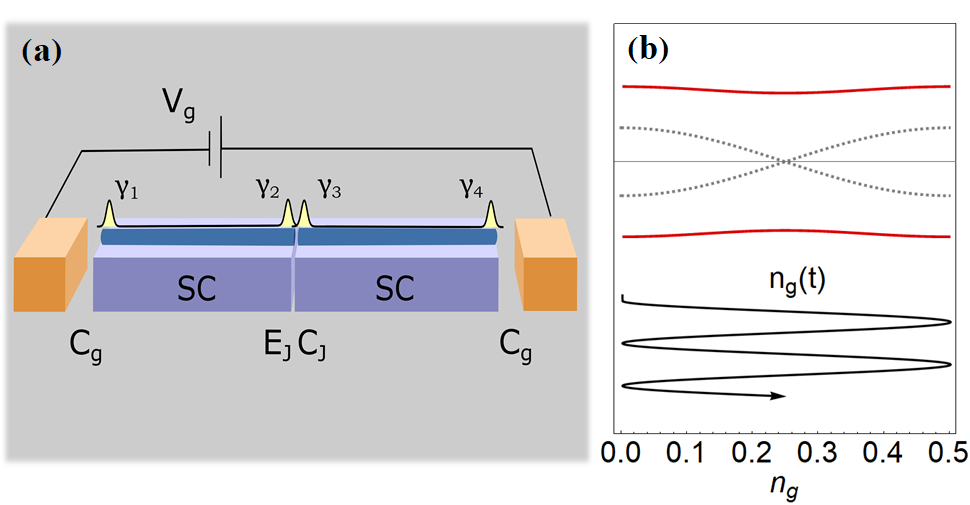}
\caption{\label{Fig1scheme} Overview of the Majorana Transmon system (MT). (a) Schematic set-up of the MT, with the two topological nanowires hosting Majorana zero modes (MZMs) $\gamma_1$-$\gamma_4$ placed on top of the leads that form the Josephson junction in a charge qubit. In this way, the two topological superconductors present the charge qubit Hamiltonian as bulk Hamiltonian. An interaction term of the two neighbouring MZMs $\gamma_2$ and $\gamma_3$ is introduced in the system and hybridises states with different fermionic parity. There is no charge transfer between the electrostatic gates at potential difference $V_g$ (coloured in orange) and the other parts of the system. (b) Schematic representation of the qubit energy dispersion and control. The qubit subspace coincides with the lowest doublet of the MT and it is represented by the red lines, while the dotted black lines are the uncoupled transmon levels. The system $H_T+H_M$ presents an avoided crossing at $n_g=1/4$ so that a sinusoidal modulation of the internal parameter $n_g$ (i.e the voltage gate $V_g$) sweeping across the range $[0,\,0.5]$ will introduce nonadiabatic transitions in the system.}
\end{figure}

\section{The Majorana transmon qubit in the instantaneous eigenbasis}
\label{sec:TheModel}
The hybrid system studied is based on the basic charge superconducting qubit where a Josephson junction (JJ) is capacitively coupled to an external voltage gate $V_g$ (also known as a Cooper pair box). In order to introduce topological states in the system, a nanowire that can host Majorana zero modes in its topological phase is placed across the JJ, so that the topological phase can be induced just in the proximity of the right and left superconducting leads and the equivalent of two topological nanowires are created on top of them. The nanowire used can be either a semiconductor with strong spin-orbit coupling or a topological insulator, two materials where, in the presence of an external magnetic field, the formation of these topological excitations at their edges has been predicted \cite{Fu2008,Lutchyn2010, Oreg2010, Cook2011}. Figure \ref{Fig1scheme}a presents a scheme of the hybrid system hosting the localised Majorana zero modes, represented at the edges of the topological nanowires. Besides the energy contributions $H_T$ coming from the superconducting part, the model includes the tunnelling interaction term $H_M$ originated from the partial overlap of the neighbouring MZMs near the Josephson junction. This term can be tuned by a local voltage gate at the Josephson junction \cite{Ginossar2014} and hybridises the eigenstates $\ket{\Psi_k^{e/o}}$ of $H_T$ with different relative parity, that is the parity of the relative number of fermions across the Josephson junction. We work in the transmon regime $E_J \gg E_C$, with $E_J$ and $E_C$ the Josephson and charging energies of the system, respectively, where the eigenstates of $H_T$ can be approximated by the eigenstates of the harmonic oscillator at the zeroth-order perturbation theory \cite{Koch2007}. In this regime, when the Majorana interaction energy $E_M$ is small compared to the charging energy, $E_M \ll E_C$, the term hybridises only eigenstates of $H_T$ within the same energy band $k$, creating a doublet structure in the spectrum. The total Hamiltonian $H=H_T + H_M$ can be written into a block diagonal form in the basis of $\ket{\Psi_k^{e/o}}$, each block representing the subspace at fixed transmon band $k$ \cite{Yavilberg2015}:
\begin{equation}
\label{HMTk}
    H^{(k)} = \begin{pmatrix} \epsilon_k^{h.o.} + t_k \cos{(2\pi n_g)} & E_M \\
                                E_M             &  \epsilon_k^{h.o.} - t_k \cos{(2\pi n_g)} \end{pmatrix}.
\end{equation}
Here $\epsilon_k^{h.o.}$ is the $k$-th energy level of a harmonic oscillator with frequency $\sqrt{8E_CE_J}$ and the term $t_k\cos{(2\pi n_g)}$ is the transmon energy dispersion derived with the WKB treatment, with  $t_k\propto E_C(-1)^{k+1} (E_J/E_C)^{k/2 +3/4}e^{-\sqrt{8E_J/E_C}}$,  $k\in\mathbb{N}$. In particular, $n_g$ is a dimensionless parameter representing the induced offset charge between the superconducting islands, in units of the Cooper pair charge 2e, and can include the contribution coming from the action of an external electric voltage on the qubit, $V_g \propto n_g$ (additional details about the solutions of this model can be found in Appendix \ref{appAspectrum}). It should be noted that the scheme shown in Figure \ref{Fig1scheme} is consistent with the use of the parameters regime $E_M\ll E_C\ll E_J$. The energies $E_M$ and $E_J$ are in fact both tunnelling parameters and so they are directly proportional to the nanowire and the superconducting junction cross-sectional area, respectively. More realistic modellings of devices leading to equation (\ref{HMTk}) exist in the literature \cite{Li2018, Yavilberg2019} and support its validity in the aforementioned parameter regime. 

One of the consequences of working with two very different energy scales $E_M$ and $E_J$ is the high anharmonicity gained in the spectrum, given by $\alpha_r = (\epsilon_1^{h.o.}-\epsilon_0^{h.o.})/2E_M = \sqrt{2E_CE_J}/E_M\gg 1$. The lowest doublet remains well isolated from the rest of the spectrum and can form the qubit computational subspace. Another important feature of this model is the exponential suppression of the intradoublet dipole interaction with the electrostatic field. This makes the chosen qubit subspace robust against radiative decay, but it forces the use of higher transmon bands when it is controlled via a superconducting cavity \cite{Yavilberg2015}. Here we aim to control the qubit without exiting the computational subspace, so that the Hamiltonian takes the form
\begin{equation}
\label{Hamiltonian}
 H\equiv H^{(k=0)}= t_0 \cos{(2\pi n_g)} \sigma_z + E_M\sigma_x\,,   
\end{equation}
where we have dropped a constant term $\epsilon_0^{h.o.}$, and $t_0<0$. The solutions of the qubit Hamiltonian (\ref{Hamiltonian}) can be expressed in terms of the states at fixed parity $\ket{\Psi^{e/o}}\equiv\ket{\Psi_{k=0}^{e/o}}$ as rotated-parity states, i.e.,
\begin{eqnlist}
       \label{PsiQubit}
    &{}\ket{\Psi^+}= \sin{\eta}\ket{\Psi^{e}} + \cos{\eta}\ket{\Psi^{o}}, \\
    &{}\ket{\Psi^-}= \cos{\eta}\ket{\Psi^{e}} - \sin{\eta}\ket{\Psi^{o}}, \\
    &{}E_{\pm} =\pm \sqrt{E_M^2 + t_0^2\cos^2{(2\pi n_g)}},
\end{eqnlist}
where $\eta[n_g] \equiv \frac{1}{2}\mathrm{atan2}\left[ E_M, |t_0| \cos{(2\pi n_g)}\right]$ is the mixing angle. We notice that the system presents an avoided crossing at $n_g =1/4$ (see Figure \ref{Fig1scheme}b), thus we decide to study the effect of a nonadiabatic modulation of such internal parameter via the external gate $V_g$. We derive the effective time-dependent Hamiltonian $H_I(t)$ of the system in the eigenstates basis $\{\ket{\Psi^{+}},\ket{\Psi^{-}} \}$ when a generic modulation $n_g(t)$ is applied. We call $\ket{\psi(t)}$ and $\ket{\psi_I(t)}$ the states of the system respectively expressed in the bases $\{\ket{\Psi^{e}},\ket{\Psi^{o}} \}$ and $\{\ket{\Psi^{+}},\ket{\Psi^{-}} \}$, while $R_y[\eta(t)]\equiv e^{i\eta(t)\sigma_y}$, with $\eta(t)\equiv  \eta[n_g(t)]$ the mixing angle defined above, is the change-of-basis transformation that connects them. Inserting $\ket{\psi(t)}=R_y[\eta(t)]\ket{\psi_I(t)}$ into the Shr\"{o}dinger equation for $\ket{\psi(t)}$, the effective Hamiltonian for $\ket{\psi_I(t)}$, $H_I(t) \equiv R_y^{-1}[\eta(t)]H(t)R_y[\eta(t)]-i\hbar R_y^{-1}[\eta(t)]\partial_tR_y[\eta(t)]$, is obtained as
\begin{eqnarray}
\label{HnonAd}
    H_I(t)=&& -\sqrt{E_M^2 + t_0^2 \cos^2\left(2\pi n_g(t)\right)}\,\sigma_z \nonumber\\ &&\quad- \frac{h\,E_M t_0 \sin[2 \pi n_g(t)]}{2\left[E_M^2 + t_0^2 \cos^2\left(2\pi n_g(t)\right)\,\right]}n_g'(t)\,\sigma_y,
\end{eqnarray}
with $n_g'(t)\equiv\partial_t n_g(t)$. This demonstrates that $n_g(t)$ introduces a direct driving term between the two qubit eigenstates $\ket{\Psi^{+}}$ and $\ket{\Psi^{-}}$. To maximise the effect of this nonadiabatic modulation, we can choose a function of $n_g(t)$ that sweeps multiple times through the range $[0,1/2]$, passing by the avoided crossing at $n_g=1/4$. A condition for the sweeping speed can be calculated to reduce the leakage to higher doublets of the system. Denoting by $\ket{\Psi_n}$ and $E_n$ a generic eigenstate and eigenenergy of the Hamiltonian $H=H_T + H_M$, this condition is given by \cite{Messiah1975}:
\[ \frac{1}{n_g'(t)} \gg \max\left|\frac{[\partial H/\partial n_g]_{mn}}{(\Delta E_{mn})^2} \right| \sim \frac{1}{E_J} \left(\frac{E_J}{32E_C} \right)^{1/4},\]
where $[\partial H/\partial n_g]_{mn}=\braket{\Psi_m|\partial H /\partial n_g|\Psi_n}$ and $\Delta E_{mn} = E_{m} - E_{n}$. Because of the fact that $E_J\gg E_M$, two very different energy scales are present in the system, which are the interdoublet gap energy $\Delta\epsilon_{01}^{h.o.}=\epsilon_{1}^{h.o.}-\epsilon_{0}^{h.o.}$ and the intradoublet energy $2E_M$, with $\Delta\epsilon_{01}^{h.o.} \gg 2E_M$. Thus, a sweeping speed $n_g'(t)$ can be chosen in order to maintain the nonadiabatic transitions within the qubit subspace.

\section{The gate protocol}
\subsection{System dynamics under parametric modulation}
\label{Sec:Dynamics}
Hamiltonian (\ref{HnonAd}) describes the dynamics originating from a generic time-dependent modulation of the parameter $n_g$, using $\{\ket{\Psi^{+}},\ket{\Psi^{-}} \}$ as the computational basis. We now aim to find a configuration of the system that results in the implementation of a specific single-qubit gate $G$ after a time $t_F$. To gain more understanding of the dynamics,  we analyse $H[n_g(t)]$ in the fixed parity basis $\{\ket{\Psi^{e}},\ket{\Psi^{o}} \}$, Eq. (\ref{Hamiltonian}), and project it onto the eigenstates basis at the end of the protocol. Also, in this section the convention $\hbar=1$ is used. We choose to modulate the $n_g$ parameter as a sinusoidal function centred at the avoided crossing $n_g=1/4$ and with amplitude $1/4$:
\begin{equation}
    \label{ngOsc}
    n_g(t)= \frac{1}{4} \Big[1 - \cos\left( \omega t\right)\Big].
\end{equation}
Here $\omega$ is the frequency of the oscillation and $T=\pi/\omega$ represents half of the period, i.e., the time it takes $n_g$ to sweep the range $[0,1/2]$ once. Since the applied external voltage $V_g$ is directly proportional to $n_g$, both share the same time dependence. In particular, both start and end smoothly at zero, making the signal straightforward to implement in realistic setups (see Appendix \ref{appCexperimental}). The frequency of the signal is suitably optimised later depending on the single-qubit gate one wants to implement. Using this modulation, the time-dependent Hamiltonian of the system in the fixed parity basis $\{\ket{\Psi^{e}},\ket{\Psi^{o}} \}$ is obtained as
\begin{equation}
\label{Ham0}
H(t)= -|t_0| \sin\big[(\pi/2)\, \cos(\omega t)\big]\sigma_z +  E_M \sigma_x,    
\end{equation}
with $\sigma_{i=x,y,z}$, being the Pauli operators. In this section we make use of the Jacobi-Anger expansions \cite{abramowitz_handbook_2013},
\begin{eqnlist}
\label{JacobiAngerCos}
\begin{split}
e^{iz\cos\theta} &{}\equiv J_0(z) + 2\sum_{n=1}^\infty (-1)^n J_{2n}(z)\cos\left(2n\theta\right) \\
&{}\quad - 2i\sum_{n=1}^\infty (-1)^n J_{2n-1}(z)\cos\left[(2n-1)\theta\right],
\end{split}\\
\label{JacobiAngerSin}
\begin{split}
e^{iz\sin\theta} &{}\equiv J_0(z) + 2\sum_{n=1}^\infty J_{2n}(z)\cos\left(2n\theta\right) \\
&{}\quad + 2i\sum_{n=1}^\infty J_{2n-1}(z)\sin\left[(2n-1)\theta\right],
\end{split}
\end{eqnlist}
to find an approximation of the system's dynamics. In particular, we employ Eq.$\,$(\ref{JacobiAngerCos}) to express the sinusoidal term in Hamiltonian (\ref{Ham0}) as a series expansion in cosines, and keep the zeroth-order term:
\begin{equation}
\label{HamOscIn}
    H'= -2 |t_0|  J_1(\pi/2) \cos(\omega t)\sigma_z + E_M \sigma_x.
\end{equation}
The system is now analogous to a two-level system coupled to a sinusoidal external drive. While the ordinary rotating-wave approximation is valid only for very small values of the ratio $2t_0 J_1(\pi/2)/E_M$, we apply the counter-rotating hybridised rotating-wave (CHRW) method \cite{Lu2012} that is valid in a much larger area of the parameters' space. Instead of removing progressively the counter-rotating-wave contributions at different orders of harmonics as is done for obtaining the Bloch-Siegert Hamiltonian \cite{Klimov2009}, the CHRW method uses a transformation that contains the contributions of all the harmonics, as we now detail. Letting $A=2|t_0| J_1(\pi/2)$, we define the unitary transformation
\begin{equation}
\label{T-transform}
    \mathcal{T}= e^{-i(A/\omega)\, \kappa \sin(\omega t) \sigma_z},
\end{equation}
with $\kappa$ a free parameter that will be conveniently chosen later [see Eq.$\,$(\ref{eqCondXi})]. Again using the Jacobi-Anger expansions, Eq. (\ref{JacobiAngerSin}), the transformed Hamiltonian $\Tilde{H}= \mathcal{T} H' \mathcal{T}^\dagger + i (\partial_t\mathcal{T}) \mathcal{T}^\dagger$ takes the form
\begin{align}
\label{Htilde0}
    \Tilde{H} = &{} -A(1-\kappa)\cos(\omega t)\sigma_z + E_M \cos\left(\frac{2A\kappa}{\omega}\sin(\omega t)\right) \sigma_x \nonumber\\
    &{}+ E_M \sin\left(\frac{2A\kappa}{\omega}\sin(\omega t)\right) \sigma_y \nonumber\\
    \simeq &{} -A(1-\kappa)\cos(\omega t)\sigma_z + E_M J_0\left( \frac{2A\kappa}{\omega}\right) \sigma_x \nonumber\\
    &{}+2E_M  J_1\left( \frac{2A\kappa}{\omega} \right) \sin(\omega t) \sigma_y\,,
\end{align}
where we have neglected the higher-order harmonic terms. The coefficients of the $\sigma_z$ and $\sigma_y$ are time dependent; hence, it is useful to work in the basis of $\sigma_x$, and express the approximated Hamiltonian (\ref{Htilde0}) in terms of its ladder operators $\sigma_+^{(x)}$ and $\sigma_-^{(x)}$, with $\sigma_{+,-}^{(x)} = \frac{1}{2}\left(\sigma_z \mp i \sigma_y\right)$:
\begin{equation}
\label{Htilde}
\begin{split}
    &{}\Tilde{H} \simeq\ E_M J_0\left( \frac{2A}{\omega}\kappa \right) \sigma_x \\
    &{}- \left\{ \frac{A}{2}(1-\kappa) - E_M  J_1\left( \frac{2A}{\omega}\kappa \right)\right\} \left(  e^{i\omega t}\sigma_+^{(x)}\ + e^{-i\omega t}\sigma_-^{(x)} \right)\\
    &{}- \left\{ \frac{A}{2}(1-\kappa) + E_M  J_1\left( \frac{2A}{\omega}\kappa \right)\right\} \left(  e^{i\omega t}\sigma_-^{(x)}\ + e^{-i\omega t}\sigma_+^{(x)} \right).
\end{split}
\end{equation}
We choose the variable $\kappa$ such that the counter-rotating terms $e^{i\omega t}\sigma_+^{(x)}$ and $e^{-i\omega t}\sigma_-^{(x)}$ of Eq. (\ref{Htilde}) vanish:
\begin{equation}
\label{eqCondXi}
    A(1-\kappa) - 2E_M J_1\Big( \frac{2A}{\omega}\kappa\Big) = 0.
\end{equation}
In this way, after neglecting the higher-order harmonic terms, the CHRW-approximated Hamiltonian takes the form
\begin{equation}
\begin{split}
    \tilde{H}_{\text{CHRW}} \simeq&{} \ E_M J_0(z) \sigma_x \\
    &{} - 2 E_M J_1(z) \left( e^{-i\omega t}\sigma_+^{(x)} + e^{+i\omega t}\sigma_-^{(x)} \right),
\end{split}
 \end{equation}
where we defined $z \equiv 2A \kappa /\omega$. We set $\Delta = 2E_M J_0(z)$ and $g= 8 E_M J_1(z)$, and we express $\tilde{H}_{\text{CHRW}}$ in the rotating frame of the drive as
\begin{equation}
\label{HchrwRF}
\begin{split}
    \tilde{H}_{\text{CHRW}}^{(\text{RF})} &{}= V\tilde{H}_{\text{CHRW}}V^\dagger + i(\partial_tV)V^\dagger \\
    &{}= \frac{\Delta-\omega}{2} \sigma_x  - \frac{g}{4} \left( \sigma_+^{(x)}  + \sigma_-^{(x)}  \right),
    \end{split}
 \end{equation}
 where $V= e^{+i(\omega/2)t \sigma_x}$. Eq. (\ref{HchrwRF}) is equivalent to the Hamiltonian of a spin placed in a constant magnetic field \cite{Klimov2009}, so that it yields the time-evolution operator
\begin{equation}
\label{UchrwRF}
\begin{split}
    &{}\tilde{U}_{\text{CHRW}}^{(\text{RF})} = \cos\left(\frac{\Omega}{2}t\right)\mathbb{I} - i\sin\left(\frac{\Omega}{2}t\right)\frac{2\tilde{H}_{\text{CHRW}}^{(\text{RF})}}{\Omega} \\
    &{}\quad= \cos\left(\frac{\Omega}{2}t\right)\mathbb{I} + i \sin\left(\frac{\Omega}{2}t\right) \left\{ \frac{g}{2\Omega}  \sigma_z - i\frac{(\Delta-\omega)}{\Omega} \sigma_x   \right\},
\end{split}
\end{equation}
with $\Omega= \sqrt{(\Delta-\omega)^2 + g^2/4}$. In order to express the dynamics in the instantaneous eigenstates basis, we transform the evolution operator back into the laboratory frame, $\tilde{U}_{\text{CHRW}}=\, e^{-i(\omega/2)t \sigma_x}\,\tilde{U}_{\text{CHRW}}^{(\text{RF})}$, we add the phase accumulated from the first transformation and we rotate the evolution into the eigenstates basis $\{\ket{\Psi^{+}},\ket{\Psi^{-}} \}$ using $R_y[\eta(t)]= e^{i\eta(t)\sigma_y}$, with $\eta(t) = \frac{1}{2}\mathrm{atan}2\left( E_M,|t_0| \sin[(\pi/2)\, \cos(\omega t)]\right)$, as defined in Sec. \ref{sec:TheModel}. In this way, the time evolution operator in the instantaneous eigenstates' basis takes the form:
\begin{equation}
\label{eq:evolCHRW_t}
    U_I(t) \simeq R_y^{-1}[\eta(t)]\mathcal{T}^\dagger V^\dagger\tilde{U}_{\text{CHRW}}^{(\text{RF})}(t)R_y[\eta(0)].
\end{equation}
We can now see what the evolution looks like after $n$ oscillations, i.e. for $\omega t_F=n\cdot 2\pi$ or equivalently $t_F=2nT$ ($T$ was defined earlier as half of the period of oscillation). Because of the periodicity of $\mathcal{T}$ and $V$, these two operators become proportional to the identity operator $\mathbb{I}$ at $t=2nT$, explicitly $\mathcal{T}^\dagger(2nT) = \mathbb{I}$ and $V^\dagger(2nT) = (-1)^n \mathbb{I}$. For the same reason we have that $\eta(2nT)=\eta(0)\equiv \eta_0$. This leads to the following expression for the evolution operator:
\begin{equation}
\label{Uchrw2nT}
    U_I(2nT) \simeq (-1)^n R_y^{-1}[\eta_0] \tilde{U}_{\text{CHRW}}^{(\text{RF})}(2nT) R_y[\eta_0].
\end{equation}
\begin{figure}[b]
\includegraphics[width=\linewidth]{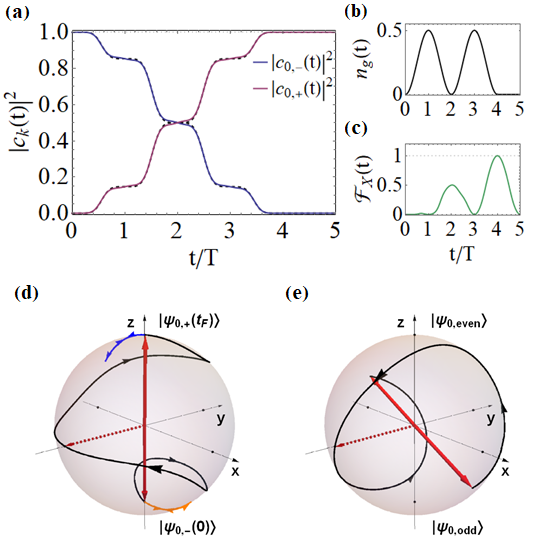}
\caption{\label{Fig2xgate} Implementation of a $n_g$-modulated X gate in the Majorana transmon system, when $\mathcal{F}_X$ is maximised for $t_F=4T$. (a) Evolution of the populations for the lowest two states of the Majorana transmon as a function of $t/T$, with $T=\pi/\omega$, using the protocol described in Sec. \ref{Sec:Dynamics}. The Black dashed lines indicate the quantity calculated without CHRW approximation. (b) Plot of the $n_g$ signal sent to the qubit. (c) Fidelity $\frac{1}{4}|tr(U_I^\dagger X)|^2$ during the operation, with its maximal value at $t=4T$. (d) Evolution of the state on the Bloch sphere, showing also the evolution of the instantaneous eigenstates in the parity basis represented in blue and orange. (e) Evolution of the state in the parity basis of $\ket{\Psi^e}$ and $\ket{\Psi^o}$. For all panels, $E_C/h=0.4$ GHz, $E_J=10E_C$ and $E_M=0.012 E_C$.}
\end{figure}
As $\tilde{U}_{\text{CHRW}}^{(\text{RF})}(2nT)$ contains only $\sigma_z$ and $\sigma_x$ terms, and as the rotation $R_y$ about the $y$-axis operates within the $xz$-plane, $U_I(2nT)$ represents a rotation about an axis that lies in this said plane. This means that potentially, for specific values of the parameters of the system, this protocol can generate single-qubit gates that include any $X$ or $Z$ rotations. To see how this protocol can represent a specific gate operation G, the fidelity $\mathcal{F}_\text{G} = \frac{1}{4}\left|\text{Tr}(U_I^\dagger(t) G)\right|^2$ can be computed. For the $\mathcal{F}_X$ we get:
\begin{equation}
\begin{split}
\label{fidX4T}
   &{} \mathcal{F}_X(2nT) \simeq\frac{\big|(\Delta -\omega)|t_0|- E_M g/2\big|^2}{\left(E_M^2 + t_0^2\right)\left(g^2/4 + (\Delta-\omega)^2\right)}\sin^2\left(n\, \Omega T\right).
\end{split}
\end{equation}
\begin{figure*}[t]
\includegraphics[width=15cm]{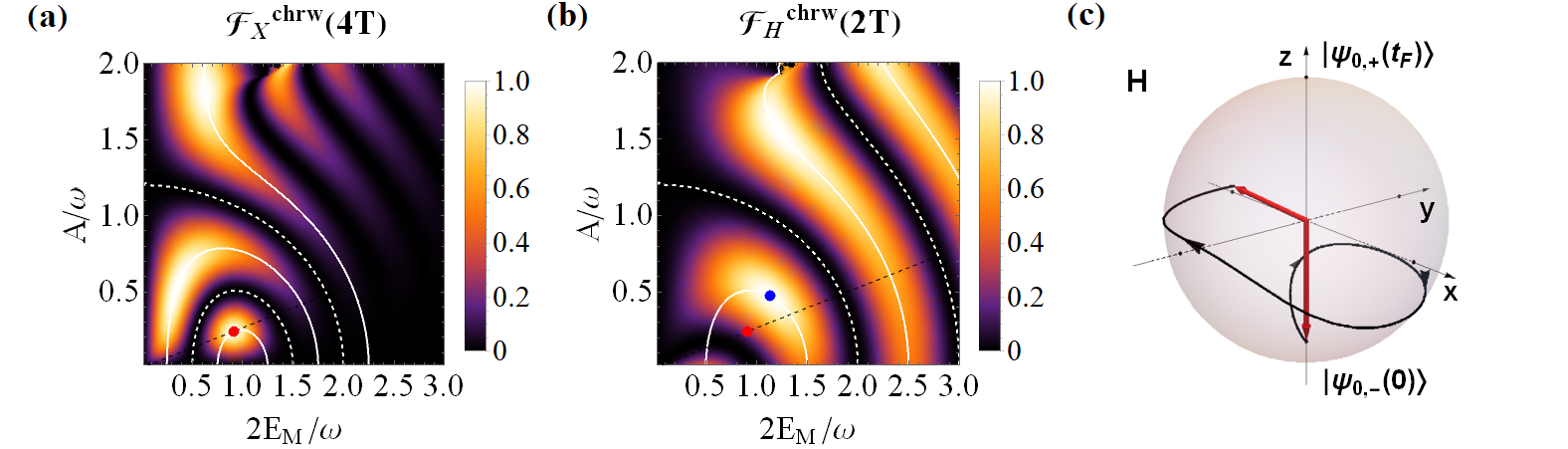}
\caption{\label{Fig3Hgate} Optimisation graphs and Bloch sphere evolution in the laboratory frame for the $X$ and hadamard $H$ gates with protocol time $t_F=4T$ and $t_F=2T$, respectively. (a) Optimisation graph representing the Fidelity for the $X$ gate as a function of $2E_M/\omega$ and $A/\omega$, with $A=2|t_0|J_1(\pi/2)$. The red dot indicates the optimal point used for the evolution in Fig. \ref{Fig2xgate}. (b)  Optimisation graph representing the fidelity for the Hadamard gate as a function of $2E_M/\omega$ and $A/\omega$, with $A=2|t_0|J_1(\pi/2)$. The blue dot indicates the optimal point of $\mathcal{F}_H$ in the laboratory frame. The optimal point for $\mathcal{F}_X$ is kept in red for comparison. In both (a) and (b) the calculation is done using the CHRW approximation. The white lines represent the resonances and antiresonances of the sinusoidal component of the two fidelities: $\sin{\left(n\, \Omega T\right)}=1$ (solid lines) and $\sin{\left(n\, \Omega T\right)}=0$ (dashed lines). The black dotted line represents the curve along which the gate can be optimised by only changing the frequency, keeping $E_J$ and $E_M$ fixed at the values optimised for the $X$ gate, in case the switch between the two gates is needed. (c) Evolution on the Bloch sphere of a state starting in the ground state $\ket{\Psi^-}$ using the optimal parameters for the Hadamard gate [blue dot in (b)]. For all panels, the charging and Josephson energies are $E_C/h =0.4$ GHz and $E_J=10E_C$.}
\end{figure*}
\begin{table*}[t]
\begin{ruledtabular}
\begin{tabular}{ p{2cm}p{0.8cm}p{1.5cm}p{1.3cm}p{1.5cm}p{1.5cm}p{1.3cm}p{1.5cm}  }
Gate & &\multicolumn{3}{c}{with CHRW} & \multicolumn{3}{c}{without CHRW} \\
 \hline
  & $t_F$ & $\mathcal{F}_{Gate}(t_F)$ & $E_M/E_C$ & $T_{Gate}$ & $\mathcal{F}_{Gate}(t_F)$ & $E_M/E_C$ & $T_{Gate}$\\
 \hline
 $X$	 & 	$4T$	 & 	$1.$	 & 	$0.012$	 & 	$189.3$ ns	 & 	$0.9998$	 & 	$0.012$	 & 	$189.8$ ns\\ 
$X$	 & 	$2T$	 & 	$1.$	 & 	$0.005$	 & 	$167.9$ ns	 & 	$0.9997$	 & 	$0.005$	 & 	$169.4$ ns  \\ 
$Hadamard$	 & 	$2T$	 & 	$1.$	 & 	$0.00759$	 & 	$186.5$ ns	 & 	$0.9998$	 & 	$0.00779$	 & 	$184.4$ ns	 \\ 
\end{tabular}

\end{ruledtabular}
   \caption{\label{tab:OptgateFixedEJ} Table of optimised values of $E_M/E_C$ and $T$ for different gates and $t_F$ at fixed $E_J/Ec=10$, in the laboratory frame. The $Z$ and $\pi/8$ are not indicated here because they can be obtained by exploiting the qubit free evolution. For all calculation, $E_C/h = 0.4$ GHz. The parameters values are suitably rounded to ensure a precision at the sixth digit for the unit fidelities (CHRW) and at the fourth digit for all the other values of $\mathcal{F}$. Some of these optimal values are indicated in the optimisation plots of Fig. \ref{Fig3Hgate}: with a red dot for $X$ gate with $t_F=4T$ and a blue dot for the Hadamard gate with $t_F=2T$.}
\end{table*}
\begin{table*}[t]
\begin{ruledtabular}
\begin{tabular}{ p{2cm}p{0.8cm}p{1.5cm}p{1.3cm}p{1.5cm}p{1.5cm}p{1.3cm}p{1.5cm}  }
\multicolumn{2}{c}{Gate (Rotating Frame)} &\multicolumn{3}{c}{with CHRW} & \multicolumn{3}{c}{without CHRW} \\
 \hline
  & $t_F$ & $\mathcal{F}_{\text{Gate}}^{\text{RF}}(t_F)$ & $E_M/E_C$ & $T_{\text{Gate}}$ & $\mathcal{F}_{\text{Gate}}^{\text{RF}}(t_F)$ & $E_M/E_C$ & $T_{\text{Gate}}$\\
 \hline
$X$	 & 	$4T$	 & 	$0.99995$	 & 	$0.012$	 & 	$189.0$ ns	 & 	$0.99997$	 & 	$0.012$	 & 	$189.5$ ns	 \\ 
$\sqrt{X}^\dagger$	 & 	$2T$	 & 	$0.99997$	 & 	$0.012$	 & 	$94.5$ ns	 & 	$0.99992$	 & 	$0.012$	 & 	$95.1$ ns 	 \\ 
$Z$	 & 	$4T$	 & 	$1.$	 & 	$0.00626$	 & 	$223.7$ ns	 & 	$1.$	 & 	$0.00626$	 & 	$223.7$ ns 	 \\ 
$Hadamard$	 & 	$6T$	 & 	$0.963$	 & 	$0.0063$	 & 	$596.0$ ns	 & 	$0.954$	 & 	$0.00627$	 & 	$599.2$ ns	 	 \\ 
$Hadamard$	 & 	$8T$	 & 	$0.982$	 & 	$0.00944$	 & 	$571.4$ ns	 & 	$0.978$	 & 	$0.00941$	 & 	$573.6$ ns	 \\ 
$T_{\pi/8}^\dagger$	 & 	$4T$	 & 	$1.$	 & 	$0.00898$	 & 	$340.0$ ns	 & 	$1.$	 & 	$0.0089$	 & 	$342.2$ ns	 	 \\
\end{tabular}

\end{ruledtabular}
   \caption{\label{tab:OptgateRF} Table of optimised values of $E_M/E_C$ and $T$ for different gates and $t_F$ in the rotating frame at the qubit frequency. For all calculations, $E_C/h = 0.4$ GHz and $E_J/E_C=10$. The parameters values are suitably rounded to ensure a precision at the sixth digit for the unit fidelities (CHRW) and at the shown significant digit for all the other values of $\mathcal{F}$. For the optimisation plots and the evolution in the Bloch sphere, see Appendix \ref{appBfidelities}.}
\end{table*}
This expression can be used to find the optimal values for $T=\pi/\omega$, $E_M$ and $E_J$ (represented by $|t_0|$). Also, specifically for the $X$ and the Hadamard gate exact conditions for these parameters can be found (see Appendix \ref{appBfidelities} for details).  As an example, Fig.$\,$\ref{Fig2xgate} shows the implementation of the $X$ gate when Eq. (\ref{fidX4T}) is maximised for $n=2$, i.e. to obtain a $X$ gate after two oscillations ($t_F=4T$). For the simulation we use values of charging and Josephson energies $E_C$ and $E_J$ that are realistic for a superconducting circuit apparatus. The resulting frequency of oscillation for $n_g(t)$ is in the order of tens of $\mathrm{MHz}$, which should be experimentally achievable given that pulses at much higher frequencies have already been implemented in the past on the first charge qubits \cite{Sillampaa2006}. A more detailed argument about the experimental implementation of this protocol can be found in Appendix \ref{appCexperimental}. In Fig.$\,$\ref{Fig2xgate}a it can be seen that the evolution of the qubit level populations under CHRW approximation, extracted from Eq.$\,$(\ref{eq:evolCHRW_t}), matches the exact evolution quite accurately. Fig.$\,$\ref{Fig2xgate}b presents the shape of the $n_g$ signal that is sent for implementing the gate, while in Fig.$\,$\ref{Fig2xgate}c the fidelity $\mathcal{F}_\text{G} = \frac{1}{4}\left|\text{Tr}(U_I^\dagger(t) G)\right|^2$ is plotted, clearly showing its maximal value at $t=4T$. In addition to this estimates, the evolution of the state of the system and the eigenstates at each value of $t\in[0,4T]$ are shown on the Bloch sphere in Figures \ref{Fig2xgate}d and \ref{Fig2xgate}e. Since we start and end the protocol at integer numbers of $n_g$ oscillations, we can notice that this specific $X$ gate with $t_F=4T$ already includes its half rotation $\sqrt{X}^\dagger$ at its midway point $t=2T$ (dashed red arrow on the same Bloch sphere), a single-qubit gate operation that can be useful when implementing a quantum algorithm.

\subsection{Other single-qubit gates and Initialisation}
\label{sec:OtherGates}
 Using the same $n_g$ modulation, Eq. (\ref{ngOsc}), other gates involving $X$ or $Z$ rotations can be obtained. Since we are working in the laboratory frame, any rotation about the $z$ axis is efficiently implemented during a free evolution. Hence we focus on other types of single-qubit gates useful for quantum computing, which are the $X$ and the Hadamard gates. Figure \ref{Fig3Hgate} compares the optimisation plots for the $X$ gate with $t_F=4T$ and the Hadamard gate with $t_F=2T$, and shows the implementation of the Hadamard gate on the Bloch sphere starting from the ground state $\ket{\Psi^-}$. Table \ref{tab:OptgateFixedEJ} shows the optimal values for these gates with different $t_F$. To obtain these values, we fix $E_J/E_C=10$ for all the calculations. In fact, if we expand $\kappa$, Eq. (\ref{eqCondXi}), up to the second order in $|t_0|$ \cite{Lu2012},
$$ \kappa \simeq \frac{\omega}{\omega + 2E_M}\left( 1 + \frac{4J_1^2(\pi/2) E_M|t_0|^2}{(\omega +2E_M)^3}\right), $$
we can use it to express the quantity $\Omega/\omega$ in terms of $2E_M/\omega$ and $|t_0|/\omega$ as
\begin{align}
    \frac{\Omega}{\omega} \simeq&{} \bigg[ \left(\frac{2 E_M}{\omega} - 1\right)^2 + \frac{8 J_1^2(\pi/2)}{2E_M/\omega + 1}\frac{|t_0|^2}{\omega^2}\frac{2E_M}{\omega} \nonumber\\
    &{}- \frac{8 J_1^4(\pi/2)}{(2E_M/\omega + 1)^3}\frac{|t_0|^4}{\omega^4}\frac{2E_M}{\omega} \bigg]^{1/2},
\end{align}
and the optimal points for the fidelity lie along the resonant curves of fixed $\Omega/\omega$ (solid, white lines in Fig.$\,$\ref{Fig3Hgate}a-b). Since $t_0(E_C, E_J)$ exponentially decreases as a function of $E_J/E_C$, the value $E_J/E_C = 10$ is chosen to balance between having short gate times (high values of the frequency of oscillation $\omega$) and remaining in the transmon regime $E_J\gg E_C$.

Either the $X$ or the Hadamard gate can be chosen for implementing quantum algorithms in the system. In fact, given that we can obtain arbitrary $z$-axis rotations under free evolution, two possible finite single-qubit gate sets \cite{Nielsen2000, BravyiKitaev2005} are the standard set of $\{H,S,T\}$ or the set consisting of the Pauli matrices and the $\pi/8$ gate, $\{X,Y,Z,T\}$, with the Pauli $Y$ gate generated using the composition of the other two, $Y=-iXZ$. Working in the laboratory frame has the advantage of not having to deal with internal parameter switching when changing the gates, because the parameters $E_J$ and $E_M$ need to be optimised only for the $X$ (and consequently $\sqrt{X}^\dagger$) or the $H$ gate, depending on the single-qubit set chosen. On the other hand, it requires precise timing during idle times, since in between two gates or algorithms the qubit has to complete an integer number of $2\pi$ $z$ rotations. 

Alternatively, the $n_g$ protocol can be studied in the rotating frame of the qubit, where the evolution operator, Eq. (\ref{Uchrw2nT}), takes the form $U_I^{\text{RF}}(2nT)=e^{-in\,\Omega_q T \sigma_z}U_I(2nT)$, with $\Omega_q=2\sqrt{t_0^2 + E_M^2}$. It this case the same type of $n_g$ modulation can be used to produce gates representing finite $z$-rotations. Table \ref{tab:OptgateRF} shows the optimised values of the fidelity and the parameters for the gates mentioned above. Notice that the phase gate can be performed using the $\pi/8$ gate, $S=T_{\pi/8}^2$. The optimisation plots and evolution on the Bloch sphere for these gates can be found in Appendix \ref{appBfidelities}. Working in this frame surely removes the need of precise timing during idle times. However, since each gate has different optimised values for $E_M$, the tuning of this internal parameter is needed when implementing a sequence of single-qubit gates. When a gate operation $G$ is applied to the qubit,  before implementing the following $\tilde{G}$ operation, the $E_M$ parameter needs to be switched from the value optimised for $G$ to that optimised for $\tilde{G}$, keeping $n_g$ fixed at $n_g=0$. This operation has to be done necessarily adiabatically in order to avoid unwanted transitions between the qubit levels. Conditions and estimates for the required switching time are derived in Appendix \ref{appCadiabSwitch}.
When choosing the Pauli-based universal gate set in the rotating frame, both methods can lead to high fidelity gates, with maximum control errors $1-\mathcal{F}_G$ of $2\times 10^{-4}$. However, in the rotating frame of the qubit the $z$ rotations turn out to be slower than during the free evolution. In fact, using the optimised values for the $X$ gate referred in Table \ref{tab:OptgateFixedEJ} in the laboratory frame, the free evolution of the qubit with a frequency $\Omega_q=2\sqrt{t_0^2 + E_M^2}$ leads to a $Z$ gate (i.e. a $\pi$ rotation about the $z$ axis) in about $50\ \text{ns}$. In contrast, in the rotating frame, where the $Z$ gate is performed using the $n_g$ protocol, the calculated gate time after two periods (4T) results to be about $200\ \text{ns}$ (see Table \ref{tab:OptgateRF} and Appendix \ref{appBfidelities} for details). In the end, it is important to comment on the possible initialisation process for the gate. The protocol in fact relies on the initial value of the offset charge, $n_g(0)=0$. The value of $n_g$ is hard to calibrate for $E_J \gg E_M$, but it can be easily tuned in the charging regime. Hence, a possible initialisation protocol can be to start in a regime where $E_J \sim E_C$ and $E_M>0$, calibrating $n_g(0)$ and moving the value of $E_J$ adiabatically back to the working regime.

\section{Charge noise effects in the system}
\label{sec:Noise}

In the previous section we showed how the modulation of $n_g$ can be used to control the system. Unfortunately, this also means that charge noise affecting this variable can potentially cause decoherence. This kind of noise is typical of superconducting devices and can come from different sources.  In this work we focus on the effect of the $1/f$ noise on $n_g$, which is intrinsic to the system since it is theorised to be coming from the coupling to random charge fluctuators \cite{Ithier2005}. The other intrinsic source of noise involved in the charge noise is telegraph noise due to quasiparticle poisoning that will not be considered here. In fact, even if the parity switching time is not known for systems that are theorised to carry Majorana quasiparticles, it has been measured in the range $1$-$10\ \mu$s in Josephson junction-based devices \cite{SunDiCarlo2012}, and there has been recent evidence of a parity switching time of $\sim160\ \mu$s for a semiconducting nanowire-based system \cite{Hays2018}. These values are relatively larger than the gate time of our protocol, hence we can focus on the analysis of the effect of the $1/f$ noise. Once the coupling with the noise source is assumed weak, the main effect of this dissipation channel can be considered to be a classical stochastic fluctuation of the parameter involved and can be characterised through the power spectral density (PSD) of the noise process \cite{Krantz2019,Ithier2005}. 
Stochastic noise modifies the dynamics of a two-level system depending on which component of the Bloch-sphere vector it is affecting. When the Hamiltonian presents the noise fluctuations in the $\sigma_x$ or $\sigma_y$ term, the dynamics is affected along the $z$ axis of the Bloch-sphere and leads to relaxation. On the other hand, fluctuations in the $\sigma_z$ term of the Hamiltonian affects the dynamics on the $x$-$y$ plane, leading to pure dephasing. These dissipation processes manifest themselves in the dynamics of the density matrix of the system. In particular, when the noise in the Hamiltonian is \commas{well behaved} (i.e., short correlated, with no singularity in the spectrum), as it is for white noise, either the Bloch-Redfield theory \cite{Bloch1957,Redfield1957}, the Born-Markov master equation approach \cite{CohenTannoudji_atomphoton} or the weak-damping path integral approach \cite{Weiss1999, Makhlin2004a, Makhlin2004} can be used to model the dynamics and lead to an exponential decay for both the energy levels' populations and the density matrix's coherence terms. The $1/f$ noise is usually introduced as a longitudinal fluctuation in the Hamiltonian \cite{Ithier2005, Krantz2019}. However, in some works transverse contributions have also been considered, and due to the fact that relaxation is a resonant phenomenon, a perturbative (diagrammatic) technique seems to lead, for the transverse component of the $1/f$ noise, to the same result as the Bloch-Redfield approach \cite{Shnirman2002, Wilhelm2007}. For a general, two-level system under free evolution starting from $\psi(t=0)=[c_0, c_1]^T$, assuming that the terms containing correlations between transverse and longitudinal components of the noise can be neglected and in the limit of zero temperature, the time-dependent density matrix of $\psi(t)$ under noise fluctuations takes the form
\begin{align}
&{}\rho_{\psi}(t) = \nonumber\\
&{}\begin{pmatrix} \frac{1}{2}[ 1 - (|c_1|^2-|c_0|^2)e^{-\Gamma_1 t}] &  c_0c_1^* e^{-\chi(t)} e^{-\frac{\Gamma_1}{2} t} e^{i\delta\omega t} \\ c_0^*c_1 e^{-\chi^*(t)}e^{-\frac{\Gamma_1}{2} t} e^{-i\delta\omega t}& \frac{1}{2}[1 + (|c_1|^2-|c_0|^2)e^{-\Gamma_1 t}] \end{pmatrix} \nonumber\\
\label{rhoMatrix}  
\end{align}
where the factors $e^{-\Gamma_1 t}$ and $e^{-\frac{\Gamma_1}{2} t}$, with $\Gamma_1$ relaxation rate, come from the Bloch-Redfield transversal contribution, while $f_z=e^{-\chi(t)}$ is the pure dephasing factor containing the pure dephasing rate $\Gamma_2^*$ and originates from the longitudinal contribution only. If the Hamiltonian can be written as $H=H_0 + V(t)$, with $V(t)$ containing the noise fluctuations in its parallel $V_z(t)$ and transverse $V_\perp(t)$ components, a perturbative expansion of the Liouville equation $\partial_t\tilde{\rho}(t) = -(i/\hbar)[V(t),\rho(t)]$ can be applied (see Appendix \ref{appCdephasing}). Within the assumption of weak, stationary and averaged-to-zero noise, the expressions for $\Gamma_1$ and $\chi(t)$ are given by:
\begin{equation}
\label{Gamma1genericV}
    \Gamma_1 = \frac{2}{\hbar^2} \int_{0}^{\infty} d\tau\, \braket{V_\perp(0)V_\perp(\tau)}\cos\left[E_{01}\tau/\hbar\right],
\end{equation}
\begin{equation}
    \label{XiDephgenericV}
\chi(t) = \frac{4}{\hbar^2} \int_0^{t}dt''\int_0^{t''} dt'\,\braket{V_z(t'')V_z(t')},
\end{equation}
with $\chi(t)$ dependent on the noise statistics and spectrum. In fact, once $V_\perp(t)$ and $V_z(t)$ are expressed in terms of $\delta n_g$, it can be seen that the decoherence rates are related to the autocorrelation of the noise $C_{\delta n_g}(\tau)= \braket{\delta n_g(0)\delta n_g(\tau)}$ and thus to its PSD $S_{\delta n_g}(\omega)= (1/2\pi) \int_{-\infty}^{+\infty} C_{\delta n_g}(\tau)e^{i\omega \tau}d\tau$. In particular, $\Gamma_1$ contains the value of $S_{\delta n_g}(\omega)$ at the resonant frequency of the system $E_{01}/\hbar$, while the form of $\chi(t)$ depends both on the expression of the PSD and the position of its cutoff frequencies $\omega_{\text{IR}}$ and $\omega_{\text{UV}}$ with respect to the evolution time \cite{Ithier2005}. Here we use a wide-band noise approximation, which assumes a noise bandwidth wide enough that the inverse of the dephasing time $T_2^*$ falls between the noise cutoffs, i.e. $\omega_{\text{IR}} \leq 2\pi/T_2^* \ll \omega_{\text{UV}}$. The choice of this assumption can be verified self consistently, once the values of the dephasing times are determined. Since $1/f$ noise has been detected at frequencies $f\gg 1\ \mathrm{MHz}$, as far as we obtain values of $T_2^*$ such that $T_2^*\gtrsim 1\ \mathrm{\mu s}$, we can consider the wide-band assumption valid.
An extensive description and derivation of the equations used in this section can be found in Appendix \ref{appCdephasing}.

\subsection{Free evolution} 
Specifically to our system, for the nondriven case, we can derive the relaxation and dephasing rates $\Gamma_1$ and $\Gamma_2^*$ by perturbatively expanding the Hamiltonian expressed in the diagonal basis of instantaneous eigenstates, Eq.$\,$(\ref{HnonAd}). At first leading orders of the noise contribution, the Hamiltonian is given by
\begin{align}
    \label{NonAdHamExpan0order}
    H_I&{}[n_g+\delta n_g(t)] \simeq -\sqrt{E_M^2 + t_0^2 \cos^2\left(2\pi n_g\right)}\,\sigma_z \nonumber\\
    &{}+ \frac{\pi\,t_0^2\sin\left(4\pi n_g\right)}{\sqrt{E_M^2 + t_0^2 \cos^2\left(2\pi n_g\right)}}\,\delta n_g\,\sigma_z \nonumber\\
    &{}- \frac{hE_M t_0 \sin(2 \pi n_g)}{2\left[E_M^2 + t_0^2 \cos^2\left(2\pi n_g\right)\,\right]}(\delta n_g)'\,\sigma_y + O(\delta n_g^2),\nonumber\\
\end{align}
which is in the form $H(t)=H_0\sigma_z + V_z \delta n_g\,\sigma_z + V_y (\delta n_g)'\sigma_y$, with $V_z\equiv \partial H_0/\partial n_g$, and unusually contains the derivative of the noise in the $\sigma_y$ term instead of a first-order term in $\delta n_g$. This term comes from the nonadiabatic transitions' contribution due to the fact that the noise affects the instantaneous eigenstates of the system over time. Here we therefore assume that the noise modulation is a differentiable function, and that $\partial_tn_g(t)$ can be uniquely defined. From Eq.$\,$(\ref{Gamma1genericV}), the expression of the relaxation rate can be obtained as
\begin{align}
\label{gamma1}
    \Gamma_1^{(1)} &{}\simeq \frac{2}{\hbar^2} |V_y|^2 \int_{0}^{+\infty} d\tau \braket{(\delta n_g)'(0) (\delta n_g)'(\tau)}e^{-i\omega_{01}\tau} \nonumber\\
    &{}=\frac{4\pi}{\hbar^2} |V_y|^2 S_{(\delta n_g)'}(f_{01}).
\end{align}
From the Fourier transform properties, if the PSD of the noise variable $\delta n_g$ is $S_{\delta n_g}(\omega)=\alpha/\omega$, $\omega_{\text{IR}}<\omega<\omega_{\text{UV}}$, we have that $S_{(\delta n_g)'}(\omega)= \omega^2\,S_{\delta n_g}(\omega)= \alpha\omega$. Knowing that $\omega_{01}= 2\pi f_{01}=2\pi\cdot 2\sqrt{E_M^2 + t_0^2 \cos^2\left(2\pi n_g\right)}$, the decoherence rate $\Gamma_1$ is given by
\begin{align}
\label{Gamma1extended}
    &{}\Gamma_1^{(1)}  \simeq 2(2\pi)^3 \left(\frac{\left(E_M t_0 \sin(2 \pi n_g)\right)^2}{4\left(E_M^2 + t_0^2 \cos^2\left(2\pi n_g\right)\,\right)^2}\right) 2\pi f_{01} \alpha \nonumber\\
    &{}\quad\ \ \ =\alpha(2\pi)^4 \frac{E_M^2 t_0^2 \sin^2(2 \pi n_g)}{\left[E_M^2 + t_0^2 \cos^2\left(2\pi n_g\right)\,\right]^{3/2}}.
\end{align}
For the calculation of the pure dephasing, under the assumption of wide-band $1/f$ noise, the modulus of the dephasing factor $|f_z|$ takes the form of a almost-Gaussian decay function at leading order in $1/\omega_{\text{IR}}t$ (see Appendix \ref{appCdephasing} for details), i.e.
\begin{equation}
\label{DephFactor}
    |f_z| \simeq e^{-(4/\hbar^2) |V_z|^2 \alpha \ln{\left(\frac{2\pi}{\omega_{\text{IR}}t}\right)}t^2 },
\end{equation}
from which we obtain a condition for the pure dephasing time $T_2^* \equiv 1/\Gamma_2^*$:
\begin{equation}
\label{Gamma2extended}
   \left(\frac{4\pi^2\,t_0^2\sin\left(4\pi n_g\right)}{\sqrt{E_M^2 + t_0^2 \cos^2\left(2\pi n_g\right)}} \right)^2 \alpha \ln\left(\frac{1}{f_{\text{IR}}T_2^*}\right) (T_2^*)^2 = 1.
\end{equation}
\subsection{Analysis near sweet-spots} 
From equation (\ref{NonAdHamExpan0order}) it can be seen that both the coefficients $|V_y|$ and $|V_z|$ of the Hamiltonian vanish at $n_g=0$, with $|V_z|$ vanishing also at $n_g=1/4$. To determine the decoherence effects at these sweet-spots the next leading order terms in the Hamiltonian expansion have to be taken into account.
With regards to $\Gamma_1$, this term is given by $V_y^{(2)}\delta n_g (\delta n_g)'$, with $V_y^{(2)}\equiv \partial V_y/\partial n_g$. It should be noticed that the variable representing the transverse noise contribution at this point, $\delta n_g (\delta n_g)'$, is generally not Gaussian. However, it can be shown that its spectrum is regular at $\omega=0$, so that the Bloch-Redfield approach used for the first order contribution still applies. For a stochastic, stationary process $x(t)$, we have, from the properties of the Fourier transform, $S_{x\,x'}(\omega)=(\omega^2/4) S_{x^2}(\omega)$. Hence we can determine the power spectral density for $\delta n_g(\delta n_g)'$ from the expression of $S_{\delta n_g^2}$, as follows. Knowing that, for two jointly Gaussian variables $x$ and $y$, we have that $\braket{x^2y^2}= \braket{x^2}\braket{y^2}+2\braket{xy}^2$, we can approximate $\braket{\delta n_g(0)^2\delta n_g(\tau)^2}\sim 2\braket{\delta n_g(0)\delta n_g(\tau)}^2$ so that $S_{\delta n_g^2}(\omega)$ is given by $S_{\delta n_g^2}(\omega) \sim (1/\pi)\int_{-\infty}^{+\infty}\{C_{\delta n_g}(\tau)\}^2e^{i\omega\tau}d\tau \sim 8\alpha^2\ln|\omega/\omega_{\text{IR}}|/|\omega|$ for $\omega_{\text{IR}}\ll \omega \ll \omega_{\text{UV}}$. We can see that the power spectral density for $\delta n_g (\delta n_g)'$ follows a quasilinear law $S_{\partial_t(\delta n_g^2)/2}(\omega) \sim |\omega|\ln|\omega|$.

Because of the regularity of the spectrum at $\omega\sim0$ and the short correlation time, we can assume that the results obtained from the perturbation theory in Eq.$\,$(\ref{Gamma1genericV}) are still applicable for the relaxation at the optimal point $n_g=0$ \cite{Shnirman2002}. In this way, the second-order correction to the relaxation produces an exponential decay with relaxation rate given by
\begin{equation}
\begin{split}
\Gamma_1^{(2)} &{}=\frac{4\pi}{\hbar^2} |V_y^{(2)}|^2 S_{\delta n_g (\delta n_g)'}(f_{01}) \\
&{}\sim  \frac{8\pi\alpha^2}{\hbar^2} |V_y^{(2)}|^2\, (2\pi f_{01})\ln|f_{01}/f_{\text{IR}}|,
\end{split}
\end{equation}
with $f_{01}=2\sqrt{E_M^2 + t_0^2 \cos^2\left(2\pi n_g\right)}$. With regards to $\Gamma_2^*$, the next leading term in the Hamiltonian expansion is $\frac{1}{2}V_z^{(2)}(\delta n_g)^2$, with $V_z^{(2)}\equiv \partial^2H_0/\partial n_g^2$. Because of the long-correlation time of $(\delta n_g)^2$, the Gaussian approximation and the perturbation theory are no longer valid for the determination of the pure dephasing effects at the optimal point. Instead, the Keldysh diagrams' method gives the following dynamics for the dephasing factor $f_z$ \cite{Makhlin2004} at long times:
\begin{equation}
\begin{split}
       |f_z(t)| &{}= \\
        =\mathrm{exp}&{}\left\{-\frac{t}{2}\int_{2\pi/t}^{+\infty} \ln\left(1 + \frac{4|V_z^{(2)}|^2}{\hbar^2}\left(S_{\delta n_g}(\omega)\right)^2  \right)d\omega \right\}.
\end{split}
\end{equation}
For $S_{\delta n_g}(\omega)=\alpha/\omega$ and $t\gg t_c$, with $t_c \equiv 1/(2|V_z^{(2)}|\alpha)$, this expression leads to an exponential decay, with dephasing rate given by
\begin{equation}
\label{Gamma2secondOrd}
    \Gamma_2^{*(2)}\sim - \frac{\pi}{\hbar} |V_z^{(2)}|\alpha. 
\end{equation}

\begin{figure}[t]
\includegraphics[width=\linewidth]{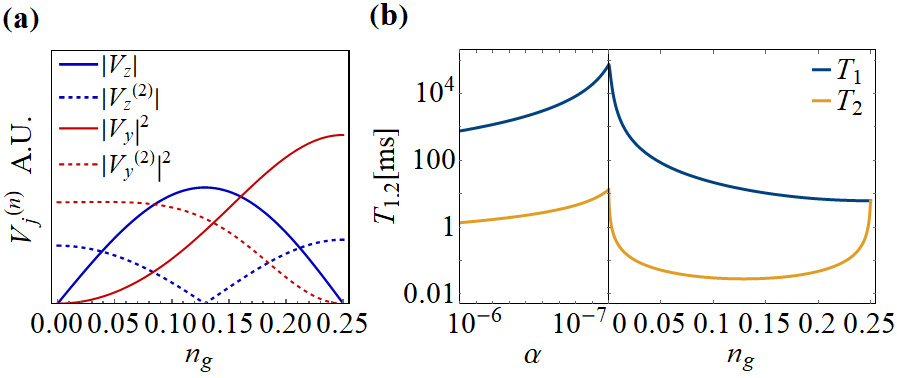}
\caption{\label{DephFixedNG} Effect of the $1/f$ charge noise on the Majorana transmon qubit free evolution. (a) Contributing elements at the first two leading orders in the expansion of the Hamiltonian with respect to the noise $\delta n_g$, see Eq.$\,$(\ref{NonAdHamExpan0order}). (b) Decoherence and dephasing times for the Majorana transmon qubit as a function of the noise strength $\alpha\in[10^{-6},10^{-7}]$ at $n_g=0$ (left panel) and as a function of $n_g\in [0, 1/4]$ for $\alpha=10^{-7}$ (right panel). For all plots, $E_C/h=0.4$ GHz, $E_J=10E_C$ and $E_M=0.012E_C$. Two sweet-spots can be notices here, with $n_g=0$ being a sweet-spot for both the relaxation and the dephasing effects.}
\end{figure}

Fig.$\,$\ref{DephFixedNG}a  and Fig.$\,$\ref{DephFixedNG}b respectively show the elements of the Hamiltonian expansion contributing to the decoherence effects, and the derived relaxation and dephasing times $T_1$ and $T_2$, with $T_1 \equiv 1/\Gamma_1$ and $T_2$ obtained by setting $\mathrm{exp}\{-\Gamma_1 T_2 /2\}f_z(T_2)=1/e$. The second-order corrections at the sweet-spot are restricted to a very narrow range of $n_g$, thus we neglect the contributions coming from cross-correlations and approximate $\Gamma_1 \simeq \Gamma_1^{(1)} + \Gamma_1^{(2)}$ and $\Gamma_2^{*} \simeq\Gamma_2^{*(1)} + \Gamma_2^{*(2)}$. The two times are plotted against the value of the parameter $n_g$ and the value of the noise strength at the sweet-spot $n_g=0$. The other parameters' values are the optimised quantities that can be used for implementing an $X$-gate operation. The noise is assumed to have $\omega_{\mathrm{IR}}=10\ \mathrm{Hz}$, which coincides with the value we use for the numerical simulation of the driven, noisy evolution described in the next section. Even if lower noise cutoffs have been reported in the literature, with values down to $\omega_{\mathrm{IR}}\sim 0.1\ \mathrm{Hz}$, 
we can see from (\ref{DephFactor}) that the dependence of the dephasing factor on $\omega_{\mathrm{IR}}$ is only logarithmic, and thus the choice of this higher value of noise cutoff carries an error which is not significant for our analysis. From Fig.$\,\,$\ref{DephFixedNG}b it can be seen that the values of the relaxation and the dephasing differ by several orders of magnitude for most of the range of $n_g$. This is due to the fact that the protocol is implemented in the basis of the eigenstates of the system, and the projection positively affects the relaxation, leaving only the contributions coming from nonadiabatic transitions between the instantaneous eigenstates. In particular, we can see that the dephasing time at the sweet-spot ranges from $1.4$ ms to $14$ ms, values which are larger than the dephasing times of the current state-of-the-art superconducting qubits \cite{Krantz2019, Burnett2019}. 
\begin{figure*}[t]
 \includegraphics[width=\linewidth]{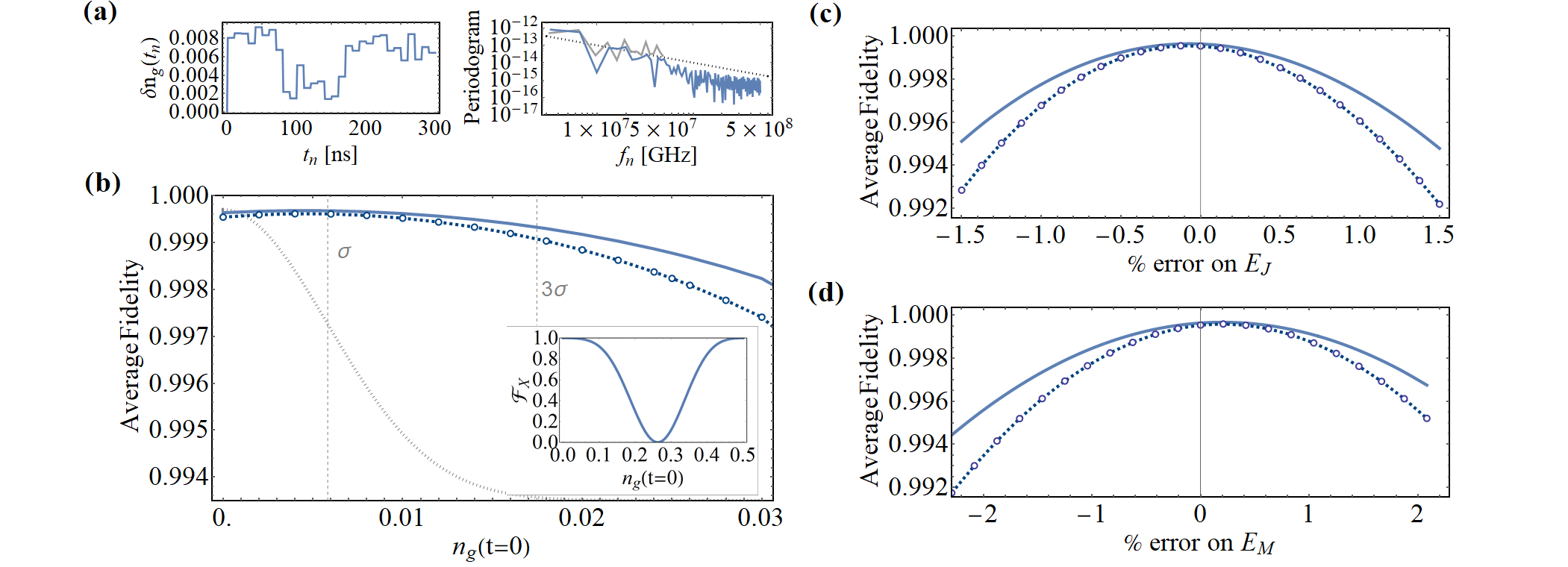}
    \caption{\label{DephDrivenStability} Effect of $1/f$ charge noise $\delta n_g(t)$ on the $X$ gate protocol with $t_F=4T$ and stability against variations of other parameters. (a) Single time-series trajectory of the simulated $\delta n_g(t)$ and related periodogram. The noise has been simulated via white noise filtering, with noise strength $\alpha\sim10^{-6}$, frequency window $10\ \text{Hz}$ - $50\ \text{MHz}$ and a time step for the stepwise constant series $\delta n_g(t_n)$ of $\delta t = 1$ ns. In the right panel, the grey curve indicates the noise jumps' periodogram while the blue curve represents the periodogram of the stepwise constant time series. (b) Fidelity of the protocol taking into account systematic error on $n_g(0)$, without charge noise (solid blue line) and with charge noise added (dash-dot blue line). The Gaussian distribution of the noise acting on the parameter is sketched in the background, with values of $\sigma_{n_g}$ and $3\sigma_{n_g}$ ($\sigma_{n_g}\sim 10^{-3}$ assuming Gaussian $1/f$ noise). Inset: fidelity of the noise-free protocol varying $n_g(0)$ from $0$ to $0.5$, showing that the protocol is quite stable near $n_g\sim0$. Also, it can be noticed that $\mathcal{F}_X$ does not change for jumps $n_g\to n_g+1/2$. (c)-(d) Fidelity of the protocol taking into account a systematic error on $E_J$ and $E_M$, without charge noise (solid blue line) and with charge noise added (dotted blue line). The parameters used at zero noise level are $E_C/h=0.4\ \mathrm{GHz}$, $E_J=10E_C$, $E_M=0.012E_C$ and $T=189.3\ \mathrm{ns}$.}
\end{figure*}

For a quantitative comparison, we can derive the relaxation and coherence times in the case of a regular transmon without the presence of the Majorana zero modes, and subjected to the same level of noise. The relaxation rate in this case can be obtained using the formula equivalent to the Fermi Golden Rule \cite{Krantz2019,YouNoriFluxQubit2007}, $\Gamma_1 = (1/\hbar^2)\left|\braket{\Psi_0|\partial H_q/\partial n_g|\Psi_1}\right|^2S_{\delta n_g}(\omega_{01})$, where $\{\ket{\Psi_0},\ket{\Psi_1}\}$ are the transmon qubit states and $\partial H_q/\partial n_g$ is the qubit susceptibility. The latter can be obtained from the expansion up to the first order in the noise fluctuation $\delta n_g$ of the transmon Hamiltonian $4E_C(n-n_g)^2-E_J\cos\left(\varphi \right)$ and yields $\partial H_q/\partial n_g \sim 8E_C n$. The charge operator $n$ has nonzero off-diagonal elements in the transmon qubit subspace. Therefore, under the harmonic approximation, we obtain \cite{Koch2007}: $\braket{\Psi_0|\left(\partial H/\partial n_g\right)|\Psi_1} \simeq 8E_C(E_J/8E_C)^{1/4}/\sqrt{2}$.  This quantity increases with $E_J/E_C$, and for the parameters $E_C=0.4\ \mathrm{GHz}$, $E_J/E_C=10$ and for a $1/f$ noise with noise strength $\alpha=10^{-7}$, it leads to a relaxation time of $T_1 = 1/\Gamma_1 \sim 0.3\ \mathrm{ms}$. Under the same harmonic approximation, this method leads to an infinite relaxation time for the Majorana transmon, since the MT intradoublet coupling $\braket{\Psi^+|\hat{n}|\Psi^-}$ vanishes \cite{Ginossar2014, Yavilberg2015}. If we compare the value of $T_1$ obtained for a regular transmon with the values shown in Fig. \ref{DephFixedNG}, we can see that the involvement of the Majorana modes improves the relaxation times by an order of magnitude ranging from $10^2$ to $10^6$ at the sweet-spot $n_g=0$. We can also see an improvement in the dephasing time. The same approach outlined above for the description of the dephasing under free evolution can be applied to a regular transmon, with the difference that in the latter case the energy terms contributing to the dephasing, $V_z$ and $V_z^{(2)}$, are proportional to the transmon energy splitting $\sim t_1-t_0$. The Majorana transmon model has a much smaller qubit splitting $\sqrt{E_M^2 + t_0^2 \cos^2\left(2\pi n_g\right)}\sim t_0$, that leads to much smaller values of $V_z$ and $V_z^{(2)}$ dictating the strength of the dephasing rate $1/T_2^*$ in Eqs. (\ref{Gamma2extended}) and (\ref{Gamma2secondOrd}). This means that, for the same values of the system and noise parameters, the coherence of the qubit is improved by a factor of $t_1/t_0 = 2^4\left( E_J/2E_C \right)^{1/2}$. Quantitatively, in the case of a transmon qubit with $E_C=0.4\ \mathrm{GHz}$, $E_J/E_C=10$ and a $1/f$ noise with noise strength $\alpha=10^{-7}$ and infrared cutoff $f_{\mathrm{IR}}=10\ \mathrm{Hz}$, the pure dephasing time ranges from $T_2^* \sim 0.3\ \mathrm{\mu s}$ at $n_g=1/4$, to $T_2^* \sim 0.15\ \mathrm{ms}$ at the sweet-spot $n_g=0$. Compared to the Majorana transmon's coherence times shown in Fig. \ref{DephFixedNG}, the improvement due to the presence of the MZMs is two orders of magnitude.

\subsection{Driven evolution}
For the driven evolution, because of the nonlinearity of the driving term used for the gate protocol, the effects of the $1/f$ noise fluctuations $\delta n_g(t)$ are studied numerically, simulating the evolution of the density operator $\rho(t)$ using a time-noise series $\{\tilde{n}_g(t_j)\}$, with $\tilde{n}_g(t_j) = n_g(t_j) + \delta n_g(t_j)$, and averaging over $10^4$ noise trajectories. To obtain a good approximation of the time evolution operator $U_I(t_N)= \prod_{j=0}^{N-1} U_I(t_{j+1},t_j)$, with $U_I(t_{j+1},t_j)=\mathrm{exp}\left\{-iH_I(t_{j+1})dt/\hbar\right\}$, we simulate a stepwise constant time series for $\delta n_g(t_j)$, with noise jumps dependent on the chosen spectral characteristics, $dt_J=1/2f_{\text{\text{UV}}}$, $f_{\text{\text{UV}}}$ being the high cutoff frequency for the noise PSD, and a smaller evolution time step $dt=1$ ns. The values of the fluctuations at each jump are calculated through a Gaussian white noise filtering in a chosen frequency range $f \in [f_{\text{IR}},f_{\text{\text{UV}}}]$. To generalise the white noise filtering including a random phase of the variable's Fourier transform, a series of complex numbers for the frequency-space $\tilde{x}_k = \frac{1}{\sqrt{2}}\left( \tilde{x}_{k,1} +i \tilde{x}_{k,2}\right)$ is generated, with $\tilde{x}_{k,1}$ and $\tilde{x}_{k,2}$ zero-average Gaussian white noise sequences with unit variance, with sampling range chosen such that the correspondent time-series step is the $dt_J$ defined above and the final time is $T_J\geq 1/f_{\text{IR}}$. The series is then filtered in the frequency space, $\tilde{y}_k= |H(f_k)|\tilde{x}_k$, using a filter amplitude of $|H(f_k)|=\sqrt{S(f_k)\,\delta f/2}$, with $S(f_k)$ sampled PSD defined for positive frequencies, $f_k=k\,\delta f$, and $\delta f =1/T_J$. Both $\tilde{x}_k$ and $|H(f_k)|$ are constructed in such a way that we obtain a real Fourier's series for the variable $\tilde{y}_k$. The resulting $1/f$ noise time-series variables used for the simulation is given by $\delta n_g(t_j)= \sum_k \tilde{y}_k e^{-i2\pi jk/N}$.  The gate fidelity used for optimising the protocol is not suitable in this case, because the evolution is no longer unitary when averaged over the noise trajectories. We need to use the more generic Uhlmann fidelity \cite{Uhlmann1976} between density operators, averaged over the Hilbert space. Luckily, the average over noise trajectories of the time-evolution operator acts as a linear, trace-preserving transformation, $\mathcal{M}[\rho(t_N)]= \braket{U_I(t_N)\rho(0) U_I^\dagger(t_N)}_{\mathrm{traj}}$. Hence, a simplified expression for the average fidelity can be used \cite{Bowdrey2002}:
\begin{equation}
    \bar{\mathcal{F}}_G  = \frac{1}{6} \sum_{j=\pm x, \pm y, \pm z} tr\left(G \rho_j G^\dagger \mathcal{M}[\rho_j] \right).
\end{equation}
Here $G$ is the target, unitary gate and $\rho_j$ represents each of the eigenstates of the Pauli operators $\sigma_j$. In fact, this expression allows us to determine $\bar{\mathcal{F}}_G$ by averaging between only some specific points in the Bloch sphere.

Figure \ref{DephDrivenStability} shows the results of the simulation for the $X$-gate protocol, with the parameters of Table I and a total evolution time of $T_G\sim 200\ \mathrm{ns}$. The noise has been produced with a PSD taking the form of $S(f)=\alpha/f$ for frequencies $10\ \mathrm{Hz} < f < 50\ \mathrm{MHz}$, and having a flat contribution $S(f)=S(10\ \mathrm{Hz})$ at lower frequencies $f\leq 10\ \mathrm{Hz}$. We also set $\alpha = 10^{-6}$. 
The use of the Fourier transform constrains us to the choice of a limited noise bandwidth. The bandwidth above, $[10\ \mathrm{Hz}, 50\ \mathrm{MHz}]$, has been chosen to ensure the inclusion of the resonant frequency of the system, of the order of $\omega_R\sim 10\ \mathrm{MHz}$. The absence of the higher cutoff in the dephasing factor (\ref{DephFactor}) can indicate that the short-time (large frequency $f \gg 1/T_G$) noise contributions are not involved in the averaged evolution (this is a consequence of both the weak coupling between the system and the noise source and the $1/f$ behaviour of the PSD), thus our choice of the higher cutoff above is based on the condition $f_{\mathrm{UV}}\gg 1/T_G\sim 5\ \mathrm{MHz}$. The chosen lower cutoff $f_{\mathrm{IR}}$ is the minimum value that we can use to ensure a high value of $f_{\mathrm{UV}}$ in the numerical simulation. Even if the actual $f_{\mathrm{IR}}$ can be lower, we can assume that the chosen value carries only a logarithmic error on the dynamics, like in the free evolution case. In the figure, one of the stepwise constant  $\delta n_g(t_j)$ trajectory is plotted in the upper panel [Fig.\ref{DephDrivenStability}a], along with its periodogram. In Figs. \ref{DephDrivenStability}b-\ref{DephDrivenStability}d the average fidelities with respect to fluctuations of the variables $n_g(t=0),\ E_J$ and $E_M$ are determined. The protocol turns out to be particularly insensitive to fluctuations of $n_g(0)$, with a fidelity reduction of $\sim0.02\%$ at $3\sigma_{n_g}$ for the combined effect of systematic and $1/f$ noise. Good results also seem to be achieved in the case of fluctuations on the other two parameters. In fact, to have a fidelity drop of $\sim0.2\%$, a systematic error of $1.5\%$ on $E_J$ or $2\%$ on $E_M$ is needed.

\section{Conclusions}
\label{sec:Conclusions}

In this work we investigated the possibility of controlling the Majorana transmon qubit, defined as the lowest doublet of the Majorana transmon \cite{Ginossar2014, Yavilberg2015} energy spectrum, exploiting a voltage-gate modulation of the induced offset charge $n_g(t)$. We modelled this dynamical modulation as a sinusoidal function such that it periodically passes through the avoided crossing point $n_g=1/4$, introducing a nonlinear driving term in the Hamiltonian, and worked in the basis of the instantaneous eigenstates of the system. Because of the high anharmonicity present in the system at $E_J \gg E_C \gg E_M$, we assumed the dynamics to be restricted to the lowest doublet of the spectrum.
We analysed the projected Hamiltonian using the counter-rotating hybridised rotating-wave method \cite{Lu2012} and we demonstrated that the effective evolution at an integer number of oscillations results in a combination of $x$ and $z$ rotations whose coefficients can be tuned using the internal and external parameters of the Hamiltonian. We then proposed two different protocols for the control of the qubit, one in the laboratory frame and the other in the rotating frame of the qubit frequency, with the first one having the advantage of faster operations and the absence of internal parameter switching, and the second one being slower but without the need of precise timing during idle times. Both the methods provide a set of single-qubit gates with control error lower than $\sim 2\times 10^{-4}$, when calculated at the zero noise level. This error is related to the limitation of the control to a simple sinusoidal function and can potentially be reduced using optimal control techniques. We also studied the effect of $1/f$ additive noise to the parameter $n_g(t)$, assuming the coupling to the noise source to be weak, and the fluctuations Gaussian, stationary, and averaged to zero. We applied a perturbative analysis to the Liouville equation and obtained an analytical expression for the relaxation and the dephasing rates under free evolution. From the calculations the system presents a sweet-spot at $n_g=0$ common to both decoherence effects. The dephasing mechanism is the one mainly affecting the system in the whole range of $n_g \in [0,0.5]$, and leads to dephasing times $T_\phi \sim 1.4-14$ ms at the sweet-spot, for noise strength in the range $\alpha\in [10^{-6},10^{-7}]$. These values for $T_\phi$ are typically beyond the current state-of-the-art transmon \cite{Cottet2002, Krantz2019}. For the driven evolution, we performed a numerical simulation of the effects of the additive noise, modelling $\delta n_g(t)$ as a stepwise constant signal with jumps produced through white noise filtering. We compared the noise-free and noisy average fidelities of the $X$ gate to derive a quantitative effect of the simulated $1/f$ noise, finding a noise-related reduction smaller than $0.01\%$. Finally, we looked at the average fidelity reduction due to systematic errors in the different parameters of the system and found low sensitivity to systematic error in the initial value of $n_g$, $n_g(0)$, and on the parameters $E_M$ and $E_J$. 

\section{Acknowledgements}
The authors acknowledge support from the European Commission's Horizon 2020 research and innovation programme under Grant Agreement No. 766714/HiTIMe. E.$\,$L. and E.$\,$Gi. gratefully thank Dr.$\ $Elinor Irish for the useful discussion about the Hamiltonian treatment and Dr$\ $Michael Stern for the useful discussion about the experimental implementation. E.$\,$Gr. acknowledges support from the Israel Science Foundation under Grant No. 1626/16.

\appendix

\section{Theoretical description of the Majorana transmon}
\label{appAspectrum}
\begin{figure}[t]
\includegraphics[width=\linewidth]{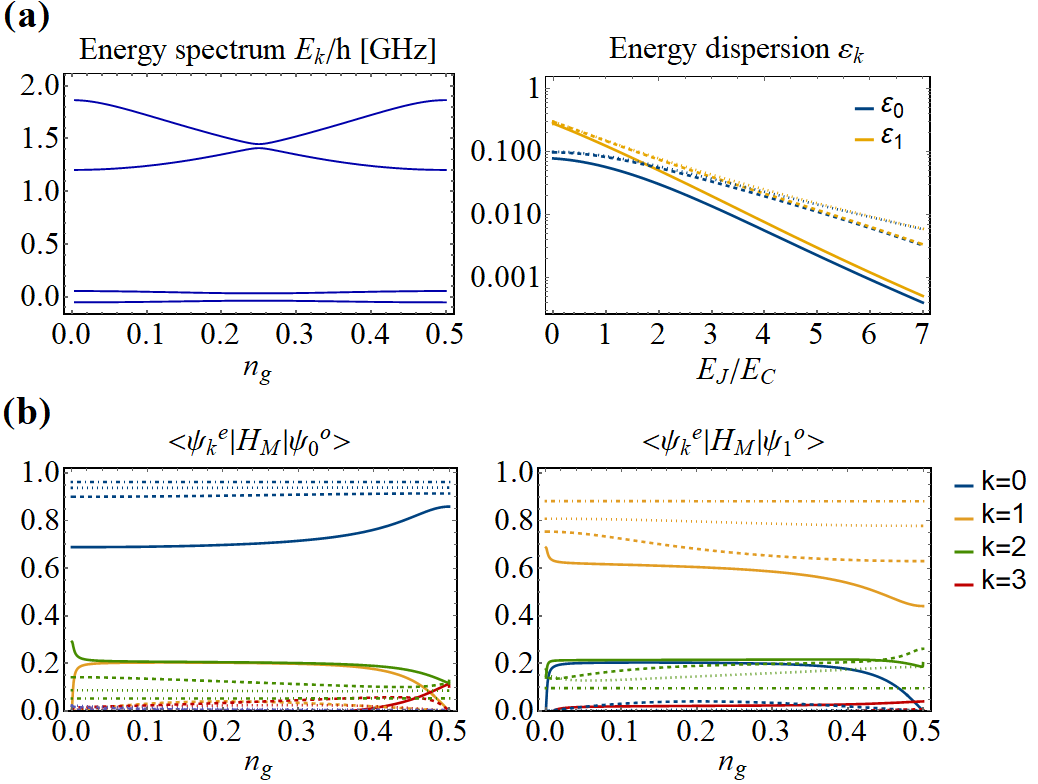}
\caption{\label{fig:MTspectrum} (a) The energy spectrum of the Majorana transmon system. \textit{left panel}: energy spectrum $E_{k,\pm}(n_g)$ as a function of $n_g$ for the first four energy levels, for $E_C/h=0.4$ GHz, $E_J/E_C=3$ and $E_M/E_C=0.1$. Even if it is not in the transmon regime, because of the Majorana interaction energy $E_M \ll E_C$, the spectrum presents a doublet-like structure with high anharmonicity. \textit{right panel}: energy dispersion $\epsilon_k =|E_{k,\pm}(0)-E_{k,\pm}(1/4)|$ for $k=0,1$ as a function of $E_J/E_C$, for different values of $E_M/E_C$: $0.12$ (solid lines), $0.012$ (dashed lines), $0.0012$ (dotted lines) (b)
Matrix elements of the interaction Hamiltonian $H_M$ (\ref{Hm}) originating from the overlap of the neighbouring Majorana zero modes near the Josephson junction and connecting transmon states of different relative fermion parity, in particular connecting the ground state (\textit{left panel}) and the first excited state (\textit{right panel}) of one of the parity sectors to the other eigenstates $\braket{\Psi_{0/1}^e|H_M|\Psi_{k}^o}$. The calculation is done in units of $E_M$, for different values of $E_J/E_C$: $0.5$ (solid line), $5$ (dashed line), $10$ (dotted line), $25$ (dot-dashed line). For all calculations, $E_C/h=0.4\ \text{GHz}$.}
\end{figure}

In this appendix we review the Majorana transmon system, which has been introduced in Ref.$\,$\cite{Ginossar2014} and applied or analysed further in a few other works \cite{Yavilberg2015, Li2018, Avila2020models, Smith2020}. We present an overview of the model for completeness of the discussion and to quantitatively support the harmonic approximation applied to the projected Hamiltonian in the lowest doublet of the spectrum, used throughout the paper. We specifically consider the model originally introduced, which describes the low-energy physics of the hybrid system in the topological phase.  As described in section \ref{sec:TheModel}, the superconducting part of the hybrid qubit consists of a traditional Cooper pair box. Its Hamiltonian, in the basis of the relative superconducting phase $\varphi=\varphi_L - \varphi_R$ between the left $L$ and right $R$ junction leads is given by
\begin{equation}
\label{Ht}
    H_T[n_g]= 4E_c \left( -i\partial_\varphi - n_g \right)^2 - E_J \cos{(\varphi)},
\end{equation}
where $ -i\partial_\varphi = \hat{n}=\frac{1}{2}(n_L - n_R)$ represents the relative number of Cooper pairs, with $n_{L}$ ($n_R$) being the Cooper pair number of the left (right) lead. The eigenfunctions $\braket{\varphi|\Psi_k}=\Psi_k(\varphi)$ of $H_T[n_g]$ are combinations of Mathieu functions \cite{Cottet2002} with boundary conditions set by the parity of the charge $n$, which is even in absence of the spare electrons, i.e. the wavefunctions have symmetric boundary conditions $\Psi_k(\varphi+2\pi)=\Psi_k(\varphi)$. When the nanowire is placed on top of the leads, the superconducting proximity effect helps the formation of the topological phase in correspondence of the two sections of the junction, and the formation of the four Majorana zero modes at their edges. With the Majorana quasiparticles being at zero energy, this setup is not sufficient to make them appear in the Hamiltonian. The Majorana transmon model takes into in account an additional interaction energy term between the neighbouring MZMs near the tunnel junction, originating from a partial overlap. This term can be modelled with a tunnelling Hamiltonian of the form
\begin{equation}
    \label{Hm}
    H_M = iE_M\cos{\left(\varphi/2\right)}\gamma_2\gamma_3,
\end{equation}
where $\gamma_2$ and $\gamma_3$ are the creation operators of the neighbouring Majorana quasiparticles, and $E_M$ represents the coupling energy. 
In terms of the electron occupation number, $H_M$ connects states of different relative parities, thus hybridising the states of the superconducting system. To see this we indicate with $N_{L,R} = 2 n_{L,R}[\text{mod}2]$ the occupation of the delocalized fermions $c_{L,R}=(1/\sqrt{2})\left(\gamma_{1,3} + i\,\gamma_{2,4} \right)$ in each nanowire, and identify the two subspaces of even and odd relative parity as \cite{Ginossar2014}
\begin{align}
         \Big\{e^{i\varphi n}\ket{N_L,N_R}:\ N_L,N_R=0,1\ \wedge\ n \in \mathbb{Z}\,\Big\}; \nonumber\\
   \Big\{e^{i\varphi n}\ket{N_L,N_R}:\ N_L,N_R=0,1\ \wedge\ n \in \mathbb{Z}+\frac{1}{2}\,\Big\}.    
\end{align}
Thus, the interaction term (\ref{Hm}) written in terms of $c_{L,R}$ and $c_{L,R}^\dagger$ allows the transitions
 \begin{equation}
     e^{i\varphi n}\ket{N_L,N_R} \leftrightarrow e^{i\varphi (n\pm 1/2)}\ket{1-N_L,1-N_R}.
 \end{equation}
The relative even/odd parity degree of freedom can be described by a two-component vector. In this way the Hamiltonian of the combined system $H_{MT}=H_T + H_M$ can be written in the form
\begin{equation}
\label{Hmt}
    H_{MT} = \begin{pmatrix} H_T[n_g] & E_M\cos{(\varphi/2)} \\
                                E_M\cos{(\varphi/2)}             &  H_T[n_g] \end{pmatrix},
\end{equation}
which is the low-energy effective Hamiltonian presented in Ref.$\,$\cite{Ginossar2014}. The delocalised fermions of the nanowires can introduce spare electrons in the system, hence the $k$th eigenfunction of $H_T$ in each parity subspace is represented by $\braket{\varphi|\Psi_k^{e}}=(\Psi_k^e(\varphi),0)^T$ and $\braket{\varphi|\Psi_k^{o}}=(0,\Psi_k^o(\varphi))^T$, with $\Psi_k^{e,o}(\varphi)$ the solution of $H_T[n_g]$ with symmetric and antisymmetric periodic boundary conditions respectively \cite{Yavilberg2015}. The hybridisation due to $H_M$ produces a doublet structure of the spectrum of the combined system. This is visible even with a low value of the ration $E_J/E_C$ (see Fig.\ref{fig:MTspectrum}). We can thus conveniently label the eigenstates of the full Hamiltonian with $\ket{\Psi_j^{\pm}}$, where $j$ is the energy band and the $\pm$ sign is related to the split levels within the band (this notation becomes even more useful when working in the transmon regime). Hence, a generic solution of Eq. (\ref{Hmt}) can be expressed as $\ket{\Psi_j^{\pm}}= \sum_k(\alpha_k^j\ket{\Psi_k^{e}}+ \beta_k^j\ket{\Psi_k^{o}})$, $\alpha_k^j$ and $\beta_k^j$ parametrically dependent on the other variables.

\begin{figure}[t]
\includegraphics[width=\linewidth]{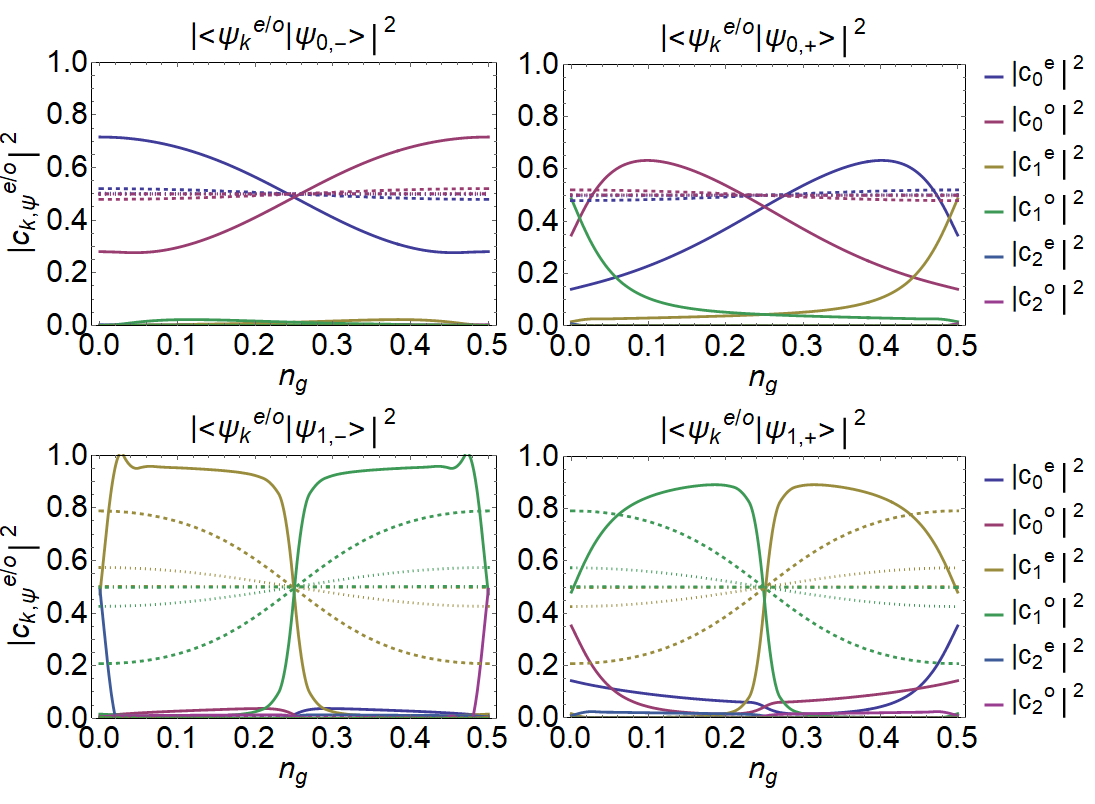}
\caption{\label{Hcomponents} Superposition of the eigenstates of the Majorana transmon with the uncoupled transmon states, each panel representing one of the first four eigenstates of the system: $\ket{\Psi_0^-},\, \ket{\Psi_0^+},\,\ket{\Psi_1^+},\,\ket{\Psi_1^-}$. The different lines correspond to contributions from different transmon energy states of different parity sectors, at different values of $E_J/E_C = 0.5$ (solid line), $5$ (dashed), $10$ (dotted), $25$ (dashed-dotted). For all calculations,  $E_M/h=E_C/h=0.4$ GHz. It can be seen that, for not too high values of $E_J$, the contribution from different parity bands is suppressed.}
\end{figure}

We are interested in the limit of high Josephson energy $E_J\gg E_C$ (transmon regime). To see the effect the interaction term $H_M$ has on the system, the energy spectrum is plotted in Fig.$\,$\ref{fig:MTspectrum}a. Even with a low value of $E_J/E_C=3$ (charging regime), the doublet structure introduced by $E_M$ is visible.
On the one hand, going towards $E_J/E_C \gg 1$ increases the anharmonicity already present at lower values, but on the other, it \commas{flattens} the energy bands. Instead, the value of the interaction energy $E_M$ is useful in changing the energy dispersion $\epsilon_k =|E_{k,\pm}(0)-E_{k,\pm}(1/4)|$ (See Fig. \ref{fig:MTspectrum}a, right panel) without affecting the anharmonicity.
 Figure \ref{fig:MTspectrum}b shows the matrix elements of the interaction Hamiltonian (\ref{Hm}) connecting the transmon eigenstates of different parities, in unit of $E_M$. Apart from a linear proportionality with respect to $E_M$, the interaction between transmon wavefunctions that belongs to different energy bands tends to zero as the superconducting qubit goes into the transmon regime. In fact, the Hamiltonian $H_T[n_g]$ in the limit of $E_J/E_C \gg 1$ resembles a anharmonic oscillator, and at zeroth order in $\sqrt{E_J/E_C}$ its eigenfunctions can be approximated by the harmonic oscillator wavefunctions. In this case, the overlap $\braket{\Psi_k^e|H_M|\Psi_l^o}$ can be shown to yield a polynomial decrease for $|k-l|$ even, and vanishing for $|k-l|$ odd, while being constant for $k=l$ \cite{Yavilberg2015}. This is even more evident when we look at the contribution of the transmon eigenstates $\ket{\Psi_k^{e/o}}$ to the linear superposition representing the eigenstates of the full Hamiltonian $H_{MT}$ in Fig. \ref{Hcomponents}. It can be seen that the hybridisation due to $H_M$ happens within each transmon band for $E_J/E_C \gtrsim 5$. Since in this work the control of the qubit is done within the doublet with $k=0$, it is reasonable to apply the harmonic approximation and neglect the interaction terms between different doublets.  Regarding the diagonalisation of $H_T[n_g]$, instead of approximating it to a anharmonic oscillator with a quartic term, we decide to use the transmon energy dispersion $\epsilon_k^T = \epsilon_k^{h.o.} \pm t_k \cos{(2\pi n_g)}$ derived from a WKB treatment \cite{Koch2007}, with $\epsilon_k^{h.o.}$ representing the $k$th eigenenergy of the harmonic oscillator with frequency $\sqrt{E_CE_J}$ and 
 $t_k \equiv (-1)^{k+1} 2^{4(k+1)}\frac{E_C}{k!}\sqrt{\frac{2}{\pi}} \left(\frac{E_J}{2E_C}\right)^{\frac{k}{2} + \frac{3}{4}}e^{-\sqrt{8E_J/E_C}}$.

\setstretch{1.01}
 The Hamiltonian assumes the block-diagonal form (\ref{HMTk}) presented in Sec. \ref{sec:TheModel}:
 \begin{equation}
\label{HMTkApp}
    H^{(k)} = \begin{pmatrix} \epsilon_k^{h.o.} + t_k \cos{(2\pi n_g)} & E_M \\
                                E_M             &  \epsilon_k^{h.o.} - t_k \cos{(2\pi n_g)} \end{pmatrix}.
\end{equation}
Each block $H^{(k)}$ can be diagonalised with a rotation about the $y$ axis, i.e.
\begin{eqnlist}
\label{solKobi}
    \ket{\Psi_k^-}=&{} \cos{(\eta_k)}\ket{\Psi_k^e} + \sin{(\eta_k)}\ket{\Psi_k^o};\\
    \ket{\Psi_k^+}=&{} - \sin{(\eta_k)}\ket{\Psi_k^e} + \cos{(\eta_k)}\ket{\Psi_k^o};\\
   E_{k, \pm} =&{}\, \epsilon_k^{h.o.} \pm (-1)^k \sqrt{E_M^2 + t_k^2\cos^2{(2\pi n_g)}},  
\end{eqnlist}
where $\pm1=s$ represents the rotated parity and $\eta_k = ((-1)^{k+1}/2)\mathrm{atan2}\left[ E_M, (-1)^{k+1}t_k \cos{(2\pi n_g)}\right]$ is the mixing angle. In this work we make use of these limit solutions, restricting the dynamics to the two-dimensional subspace of $k=0$.

\section{Experimental realisation}
\label{appCexperimental}
In this appendix we aim to explore the experimental feasibility of the $n_g$ modulation proposed for the single gate protocol. Throughout this study we use values of $E_C$ and $E_J$ that are realistic for superconducting circuit devices \cite{Krantz2019}. Arguably, small values of $E_M$ can be achieved as discussed in Sec. II. Here we want to determine the physical requirements for a clean modulation of the gate voltage of the qubit. The signal needed for the protocol,
\begin{equation}
\label{pulseAppB}
    n_g(t)=
    \begin{cases}
    \left(1 - \cos\left( \omega t\right)\right)/4 &  0\leq t\leq 2\pi n/\omega\\
    0 & \mathrm{otherwise}
    \end{cases}
\end{equation}
can be seen as a pulse with a frequency bandwidth of about tens of megahertz. We can assume that, in an hypothetical experimental apparatus, it can be reproduced by an arbitrary waveform generator (AWG), whose effect on the signal is represented by a Gaussian filter. This kind of effect has also been taken into account in works on optimal control algorithms in superconducting circuit devices that use the transfer function formalism \cite{Motzoi2011}, assuming AWG's sampling rate of 1 gigasample/sec and a Gaussian filtering attenuation of $250\ \mathrm{MHz}$ at $-3\mathrm{dB}$, meaning that it can be represented by a Gaussian filter of $300\ \mathrm{MHz}$ width. Both the values of the sampling rate and the width of the Gaussian filter are high enough to ensure a smooth interpolation of the digitalised input, and a negligible filtering effect on the output signal. A further analysis can be done regarding the physical effect of the coaxial line that is usually used for sending the voltage signal to the qubit in a superconducting circuit setup. In particular, we can find the conditions under which the impedance of the coaxial line does not alter the said voltage modulation. Considering that a transmission line has a typical characteristic (lossless) impedance $Z_0$ of $50\ \Omega$, and the impedance of a capacitor in an AC circuit is $Z_C=1/(i\omega C)$, where $C$ is its capacitance and $\omega$ is the frequency of the AC field across it, to have minimal effects coming from $Z_0$ we need all the voltage drop across the transmission line contributing to the voltage difference of the capacitance used to couple the coaxial line to the qubit, $V_C\gg V_{coax}$.
This translates to  $|Z_C| \gg |Z_{coax}|$ and thus $C\ll 1/(\omega\times50\Omega)$. Assuming that, geometrically, $C\sim\epsilon d$, with $d$ the dimension of a square capacitor, and $\epsilon\sim8.85\times10^{-10}\ \mathrm{F/m}$, it follows that, to neglect the voltage drop of the coaxial line and have an AC voltage modulation of $\sim10\ \mathrm{MHz}$ across the capacitor, the latter needs to have dimensions $d\ll 2 \times 10^3\ \mathrm{m}$, which is orders of magnitude larger than the typical size of a capacitor in a superconducting circuit ($\mathrm{\mu m}$). These arguments thus show the feasibility of a modulation of the type (\ref{pulseAppB}) in a realistic scenario, and that effects of deformation that can come from the superconducting external apparatus can be considered irrelevant at the frequency values presented in this work.

\section{Single-qubit gate fidelity expressions}
\label{appBfidelities}
\begin{figure*}[htp]
\includegraphics[width=17.2cm]{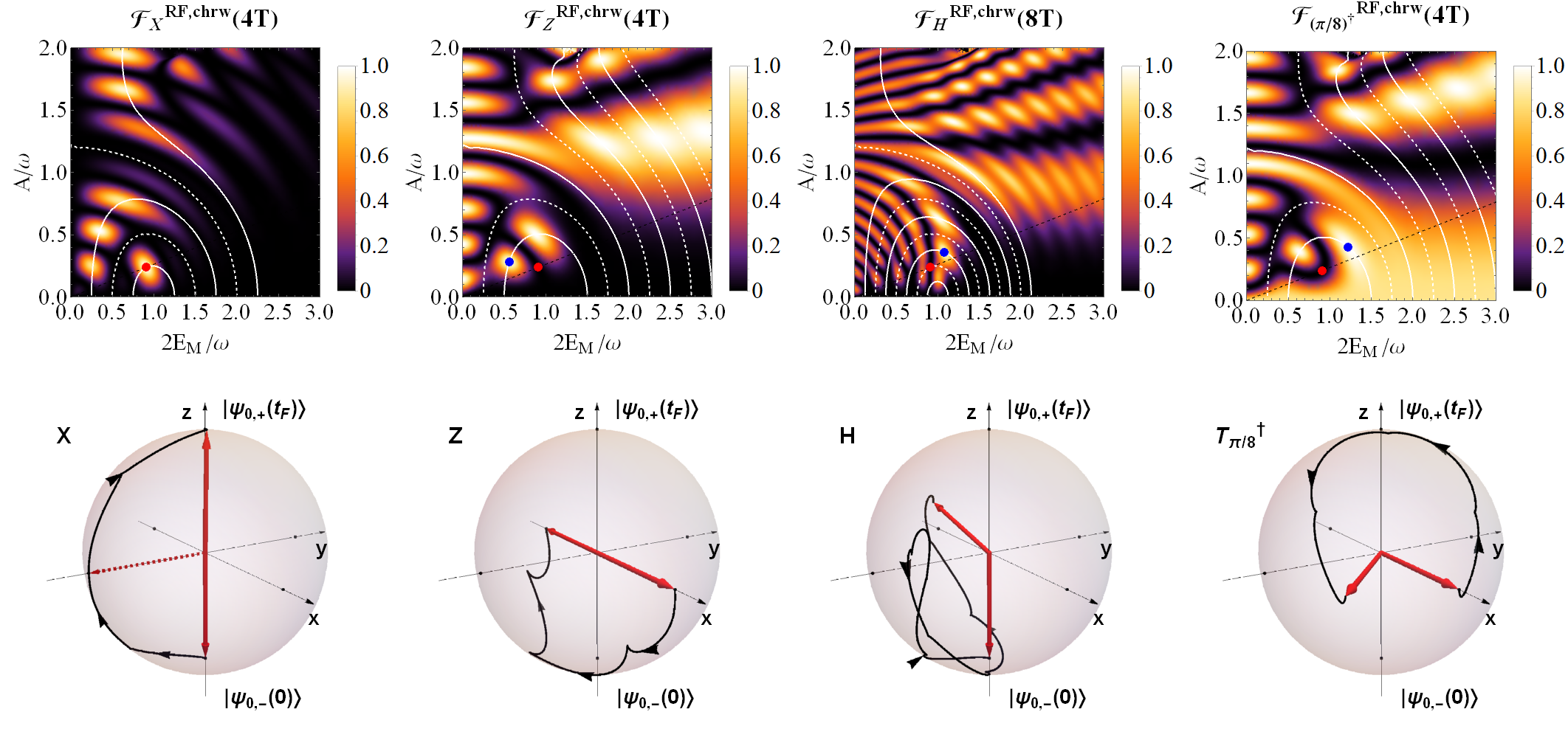}
        \caption{\label{fig:OptGateXZHTrot}
         Optimisation graphs and Bloch sphere evolution for the $X$, $Z$, Hadamard $H$ and $(T_{\pi/8})^\dagger$ gates working in the rotating frame of the qubit. Note that the $X$ gate includes $(\sqrt{X})^\dagger$ when performed halfway. \textit{Upper panels}: optimisation graphs representing the fidelity for the specified gate as a function of $2E_M/\omega$ and $A/\omega$, with $A=2|t_0|J_1(\pi/2)$.  The calculation is done using the CHRW approximation. The blue dot indicates the optimal point for the specified gate. For comparison, the red dot indicates the optimal point used for the evolution in Fig. \ref{Fig2xgate}, i.e. for the $X$ gate with protocol time $t_F=4T$. 
         In all the plots the white lines represent the resonances and antiresonances of the sinusoidal component of the fidelities: $\sin{\left(2\pi n\cdot \Omega T\right)}=0,1$ (dashed, solid lines) for $\mathcal{F}_X$ and $\mathcal{F}_H$,  $\sin{\left(2\pi n\cdot \Omega T\right)}=1,0$ (dashed, solid lines) for $\mathcal{F}_Z$ and $\mathcal{F}_{(\pi/8)^\dagger}$. The black dotted lines represent the curves along which the gate can be optimised by only changing the frequency, keeping $E_J$ and $E_M$ fixed at the values optimised for the $X$ gate. \textit{Lower panels}: each gate is implemented on a specific state, using the optimal parameters indicated in the upper panels, and its evolution in the Bloch sphere is shown. The initial states are: $\frac{1}{\sqrt{2}}\left(\ket{\Psi^+}+\ket{\Psi^-}\right)$ for the $Z$ and $(\pi/8)^\dagger$ gate, $\ket{\Psi^-}$ for the $X$ and Hadamard $H$ gate. For all panels, the charging and Josephson energy are $E_C/h =0.4$ GHz and $E_J=10E_C$.}
\end{figure*}
In this appendix we present the expressions for different single-qubit gates when using the $n_g$-modulated protocol described in section \ref{Sec:Dynamics}, working in both the laboratory and the qubit rotating frames. In particular, we present the analytical formulas for the time evolution operator in the two frames, finding exact conditions for obtaining the $X$ and the Hadamard $H$ gates in the laboratory frame and the $Z$ and the $T_{\pi/8}$ gates in the rotating frame. We also show the optimisation plots and the implementation on the Bloch sphere of some single-qubit gates in the qubit rotating frame (Fig.$\,$\ref{fig:OptGateXZHTrot}).
From Eq. (\ref{Uchrw2nT}), the evolution, in the laboratory frame, at an integer number of oscillations $t_F=2nT$ is given by
$$  U_I(2nT) \simeq (-1)^n R_y^{-1}[\eta_0] \tilde{U}_{\text{CHRW}}^{(\text{RF})}(2nT) R_y[\eta_0], $$
with $ \tilde{U}_{\text{CHRW}}^{(\text{RF})} $ given by the time-evolution expression found using the CHRW method, Eq.$\,$(\ref{UchrwRF}) in section \ref{Sec:Dynamics}, while $\eta_0 = \frac{1}{2}\mathrm{atan2}\left( E_M, |t_0|\right)$ is the mixing angle that diagonalises the qubit Hamiltonian at the beginning and the end of the evolution, Eq.$\,$(\ref{PsiQubit}). Knowing that $\cos(2\eta_0)\equiv |t_0|/\sqrt{E_M^2 +|t_0|^2}$ and $\sin(2\eta_0)\equiv E_M/\sqrt{E_M^2 +|t_0|^2}$, we can explicitly write
\begin{equation}
\label{Uchrw2nTapp1}
    U_I(2nT) \simeq \cos\left(n\, \Omega T\right) \mathbb{I} - i \frac{\sin\left(n\, \Omega T\right)}{\Omega\sqrt{E_M^2 +|t_0|^2}}\bigg(c_x \sigma_x -  c_z \sigma_z\bigg), 
\end{equation}
where we have
\begin{align*}
       \Omega=&\sqrt{(\Delta-\omega)^2 + g^2/4}, \nonumber\\
         c_z =& |t_0|g/2 + (\Delta-\omega)E_M, \nonumber\\
        c_x =&  (\Delta-\omega)|t_0| - E_M\,g/2,
\end{align*}
with the parameters given by
\begin{align*}
       \Delta =&\ 2E_M J_0\left( 2A\kappa/\omega \right), \nonumber\\
         g =&\ 8E_M J_1\left(2A\kappa/\omega \right), \nonumber\\
        \kappa= &\ \mathrm{Root}\left\{ A(1-\kappa)-2E_M J_1\left(2A\kappa/\omega\right)\right\}, \nonumber\\
        A = &\ 2|t_0| J_1(\pi/2).
\end{align*}
In general, the optimal parameters for implementing a specific gate $G$ can be determined numerically by maximising the gate fidelity $\mathcal{F}_G = \frac{1}{4}|tr(U_I^\dagger G)|^2$. However, it can be seen that for the $X$ and the Hadamard gates, which are the ones of interest when working in the laboratory frame, the parameters can be optimised exactly, looking directly at equation (\ref{Uchrw2nTapp1}): for the X gate, the parameter's conditions are
\begin{eqnlist}
\label{caseX}
 c_z=0&{} \ \ \mathrm{and}\ \ n\,\Omega T = \pi/2 + m\pi, \ \  m=0,\,1,\dots \\
 \intertext{while\ for\ the\ Hadamard\ gate,\ we \ need}
\label{caseH}
c_z=-&{}c_x \ \ \mathrm{and}\ \ n\,\Omega T  = \pi/2 + m\pi,\ \ m=0,\,1,\dots
\end{eqnlist}
As an alternative gate implementation, we want to present the dynamics of the system obtained in the rotating frame of the qubit at $t=0$. Passing to this rotating frame, the time evolution operator takes the form
\begin{align}
\label{Uchrw2nTrfQubit}
U_I^{\text{RF}}(2&{}nT)= e^{-in \Omega_q T\sigma_z}U_I(2nT) \nonumber\\
 \simeq&{}\, \cos\left(n\,\Omega_q T\right) \bigg\{\cos\left(n\,\Omega T\right) \mathbb{I} \nonumber\\
    &{}- i \frac{\sin\left(n\, \Omega T\right)}{\Omega\sqrt{E_M^2 +|t_0|^2}}\bigg( c_x \sigma_x - c_z \sigma_z \bigg) \bigg\} \nonumber\\
&{}+\, \sin\left(n\, \Omega_q T\right) \bigg\{
\frac{\sin\left(n\, \Omega T\right)}{\Omega\sqrt{E_M^2 +|t_0|^2}}\, c_z\,  \mathbb{I} \nonumber\\
&{}-i\cos\left(n\, \Omega T\right) \sigma_z 
- i \frac{\sin\left(n\, \Omega T\right)}{\Omega\sqrt{E_M^2 +|t_0|^2}}\, c_x\, \sigma_y  \bigg\},
\end{align}
where we have $\Omega_q=2\sqrt{t_0^2 + E_M^2}$. The optimisation plots for the single-qubit gates $X$, $Z$, Hadamard $H$, and $(T_{\pi/8})^\dagger$ are shown in the upper panels of Fig.$\,$\ref{fig:OptGateXZHTrot}. The $\sigma_z$ rotation of the rotating frame transformation introduces a $\sigma_y$ term in the Hamiltonian, whose coefficient vanishes when one of the terms $c_x$, $\sin\left(n\,\Omega_q T\right)$ or $\sin\left(n\,\Omega T\right)$ do. Therefore, to obtain an $X$ gate, we need  $\sin\left(n\, \Omega_q T\right)\sim0$ and the same conditions as for the laboratory frame given in Eq.$\,$(\ref{caseX}) to be fulfilled. The solutions in this case are not exact, but from the plot of the fidelity $\mathcal{F}_X^{\text{RF}}$ in Fig.$\,$\ref{fig:OptGateXZHTrot}, it can be seen that the optimised points in the area of the parameter space of interest are close to the same resonances of the laboratory frame. A similar argument can be made for the Hadamard gate, which requires both the $\sigma_x$ and $\sigma_z$ terms. Regarding the $Z$ gate, we can identify some exact points, obtained for $\cos\left(n\, \Omega_q T\right)=0$ and $\sin\left(n\,\Omega T\right)=0$. The other optimised points presumably belong to the case of $\sin\left(n\, \Omega_q T\right)=0$ but do not have an exact solution and they are not of interest for this study. In the same way a subset of exact solutions can be found for the $(T_{\pi/8})^\dagger$ gate, with the conditions $\sin\left(n\, \Omega T\right)=0$ (white lines in Fig.$\,$\ref{fig:OptGateXZHTrot}) and $n\, \Omega_q T = -\pi/8 + k\pi$, for  $k\in \mathbb{N}$. Finally, the lower panels of Fig. \ref{fig:OptGateXZHTrot} show the implementation of such gates on the Bloch sphere in the rotating frame. 
 
\section{Adiabatic condition's estimate when operating in the rotating frame}
\label{appCadiabSwitch}
In this appendix we present a realistic example of parameter switching needed when operating in the rotating frame to apply a simple sequence of single-qubit gates to the Majorana transmon qubit. In fact, when working in the rotating frame using the $n_g$-modulated protocol to implement single-qubit gates, before performing each gate operation the parameters of the system need to be tuned to their optimised values, keeping $n_g=0$. In Table \ref{tab:OptgateRF} we show how the parameters optimisation can be done by keeping $E_J$ fixed, so that only $\omega$ and $E_M$ need to be tuned when changing qubit gate. Because $E_M$ represents an internal parameter of the system, tuning its value can introduce nonadiabatic transitions between its eigenstates $\ket{\Psi^-}$ and $\ket{\Psi^+}$. Thus we need to derive the condition for having an adiabatic evolution during this operation. The switching between different values of $E_M$ can be represented by a continuous function of time $E_M(t)$ and leads to a time-dependent Hamiltonian that, in the basis $\{\ket{\Psi^-}, \ket{\Psi^+}\}$, presents a nonadiabatic $\sigma_y$ term:
\begin{equation}
    H_{Switch} = \sqrt{t_0^2 + E_M^2(t)}\, \sigma_z + \frac{|t_0|\,\partial_t E_M(t)}{4\pi\left[t_0^2 + E_M^2(t)\right]}\,\sigma_y\\.
\end{equation}
The adiabatic condition is derived imposing the $\sigma_y$ term to be negligible with respect to the diagonal $\sigma_z$ term:
\begin{equation}
\label{adcriterionEM}
\mathrm{\max_{0 \leq t \leq \tau}}\left|\frac{|t_0|\,\partial_t E_M(t)}{4\pi\left[t_0^2 + E_M^2(t)\right]^{3/2}} \right| \ll 1\,.
\end{equation}
This is also equivalent to $\mathrm{max}\left|\left[ \partial_t H(t)\right]_{mn}/(\Delta E_{mn})^2\right|\ll 1$. Condition (\ref{adcriterionEM}) needs also to be combined with the condition of obtaining an identity operation during the parameter's switch. 
To give some realistic estimates, we choose to analyse the case of the two-gate sequence $XZ$. This is an adequate case since only single-qubit gates within the chosen universal set are significant for quantum computing purposes, and we gave as example in section \ref{sec:OtherGates} the set of gates $\{H,S,T\}$ and $\{X,Y,Z,T\}$. Given that $Y=-iXZ$, we can reduce each of these sets to the use of only one gate that performs a fixed rotation perpendicular to the $z$ axis (either $H$ or $X$) and two gates performing fixed rotations about the $z$ axis. Thus, it is worth analysing the case of a sequence of these two types of single-qubit gates. When we apply the $XZ$ sequence in the rotating frame, in between these two operations we need to tune the parameter $E_M$ from $E_M^X$ to $E_M^Z$, optimised values for the $X$ and $Z$ gates, respectively. The function we choose for the smooth switch is the sinusoidal function
\begin{equation}
    E_M(t) = E_M^X + \frac{E_M^Z-E_M^X}{2}\left(1-\cos\left(\pi t/\tau\right) \right),
\end{equation}
with $\tau$ the time of the switching operation. When calculating the condition for $\tau$ for fulfilling the adiabatic criterion (\ref{adcriterionEM}), using the values of $E_M^X$, $E_M^Z$ and $t_0$ from table \ref{tab:OptgateRF}, we obtain $\tau \gg 10\ \mathrm{ns}$.
Regarding the condition for obtaining an identity operation, we take into account the fact that we are working with two different rotating frames. The frequency of the qubit is in fact changing when $0 < t \leq \tau$, so that we need to find a value for $\tau \gg 10\ \mathrm{ns}$ such that the qubit performs an integer number of $2\pi$ $z$ rotations in that time interval. The time evolution operator during this operation is represented by ($h=1$)
\begin{equation}
\label{switch1}
    U_{Switch}^{RF} = e^{i2\pi\Omega_Z\tau\sigma_z}e^{-i2\pi\int_0^\tau\sqrt{t_0^2 + E_M^2(t)}dt\,\sigma_z},
\end{equation}
where $\Omega_Z=\sqrt{t_0^2 + (E_M^Z)^2}$ is the frequency of the qubit at the end of the protocol. Since approximately $\int_0^\tau\sqrt{t_0^2 + E_M^2(t)}dt \simeq \{\sqrt{t_0^2 + E_M^2(0)}+\sqrt{t_0^2 + E_M^2(\tau)\,}\} \tau /2 = (\Omega_X + \Omega_Z)\tau/2$, we obtain the condition 
\begin{equation}
    \tau =  \left[ \frac{1}{\tau}\int_0^\tau\sqrt{t_0^2 + E_M^2(t)}dt - \Omega_Z\right]^{-1} \simeq\ 2/|\Omega_X - \Omega_Z|.
\end{equation}
In our example of $E_M^X$-to-$E_M^Z$ switch, the timescale needed is $\tau \sim 1\ \mathrm{\mu s}$. 
However, we can observe that we can think about a protocol where we switch $E_M$ from $E_M^X$ to $E_M^Z$ first, implement the $Z$ gate, and then switch the parameter back from $E_M^Z$ to $E_M^X$, using the function $\tilde{E}_M(t) = E_M^X + (E_M^Z-E_M^X)\left(1+\cos\left(\pi t/\tau\right) \right)/2$. In this case, since the propagators for each step of this protocol represent $z$ rotations and they all commute with each other, the effect of the $E_M^Z$-to-$E_M^X$ switch, 
\begin{equation}
    \tilde{U}_{Switch}^{RF} = e^{i2\pi\Omega_X\tau\sigma_z}e^{-i2\pi\int_0^\tau\sqrt{t_0^2 + \tilde{E}_M^2(t)}dt\,\sigma_z},
\end{equation}
approximately cancels the contribution from the $E_M^X$-to-$E_M^Z$ one, Eq. (\ref{switch1}), so that the propagator for the two-gate sequence becomes $U_{XZ}^{RF}(t) = \tilde{U}_{Switch}^{RF}U_Z^{RF}U_{Switch}^{RF}U_X^{RF} \simeq U_Z^{RF}U_X^{RF}$, given that condition (\ref{adcriterionEM}) is fulfilled. We can conclude that, when we want to apply the $X Z$ sequence to the Majorana transmon, we can decide to start and end the sequence in the system's configuration with $E_M = E_M^X$, so that the condition for having an identity operation during the switching of $E_M$ is always fulfilled, while the adiabaticity criterion leads to the switching time's condition $\tau\ \gg\ 10\ \mathrm{ns}$. This is also valid for any two-gate sequence that includes a finite rotation about the $z$ axis.

\section{Charge 1/f noise Decoherence rates expressions}
\label{appCdephasing}
\setstretch{1.05}
We derive the expressions for the dephasing and decoherence rates during free evolution used in Sec. \ref{sec:Noise}, in particular Eqs. (\ref{Gamma1genericV}), (\ref{XiDephgenericV}), (\ref{gamma1}) and (\ref{DephFactor}). In the literature, several works study the dissipative effects of the $1/f$ noise on the evolution of a two-level system \cite{Makhlin2004a, Shnirman2002, Ithier2005, Knapp2018}. However, the system considered in this work presents some noise contribution to the perpendicular component of the Hamiltonian, which is usually neglected. Here a more detailed review of the processes involved is desirable. We assume that the effect of the weak, stationary noise on a two-level system can be represented in the Hamiltonian by a classical stochastic perturbation of its longitudinal and transverse components. If we call $x_i(t)$ the variable along the $i$-axis where the noise component $\delta x_i(t)$ is acting, the Hamiltonian evolving under free evolution can be approximated to $H[\delta x(t)]\simeq H_0\sigma_z + V(t)$, with $V(t)$ the term containing the noise fluctuations. In the case of the Majorana transmon we have $V(t) = V_z[\delta x_z(t)]\sigma_z + V_\perp[\delta x_\perp(t)]\sigma_y \equiv V_z(t)\sigma_z + V_\perp(t)\sigma_y$. The Liouville equation in the interaction picture is $\partial_t\tilde{\rho}(t) = -(i/\hbar)[\tilde{V}(t),\tilde{\rho}(t)]$, with the density matrix given by $\tilde{\rho}(t)= e^{i(H_0t/\hbar)\sigma_z}\rho(t)e^{-i(H_0t/\hbar)\sigma_z}$, and
$\tilde{V}= e^{i(H_0t/\hbar)\sigma_z}V(t)e^{-i(H_0t/\hbar)\sigma_z}\equiv V_z\sigma_z + V_\perp\left( e^{i(\Omega_{01}t-\pi/2)}\sigma_p^z + e^{-i(\Omega_{01}t-\pi/2)}\sigma_m^z\right)$, 
where we have named $\Omega_{01}=2H_0/\hbar =E_{01}/\hbar$, with $E_{01}$ qubit energy splitting, and $\sigma_{p,m}^z=\frac{1}{2}\left(\sigma_x \pm i\sigma_y\right)$. The Liouville equation is iterated once and averaged over the noise ensemble, yielding
\begin{equation}
    \dot{\tilde{\rho}}(t) = -\frac{1}{\hbar^2} \int_0^t dt' \braket{[\tilde{V}(t), [\tilde{V}(t'), \tilde{\rho}(t')]]},
\end{equation}
where we have set the noise average to zero  $\braket{\tilde{V}(t)}=0$. This is widely assumed when calculating the decoherence rates, since every possible nonzero contribution can potentially be included into $H_0$ \cite{CohenTannoudji_atomphoton}.  Since the noise source is weakly coupled to the two-level system, we can assume that it does not significantly change the components of the density operator within the evolution timescale; hence, we can approximate $\tilde{\rho}(t') \simeq \tilde{\rho}(t)$. The expressions of the diagonal and transverse components take the form
\begin{widetext}
\begin{equation}
\label{liuvilleExpanded}
    \begin{split}
       \dot{\tilde{\rho}}_{00,11}(t)  =&{} \pm\frac{2}{\hbar^2} \int_0^{t} dt'\,\bigg\{ \braket{V_\perp(t)V_\perp(t')}\cos\left[\Omega_{01}(t-t')\right]\Big(\tilde{\rho}_{11}(t) - \tilde{\rho}_{00}(t)\Big) \\
      &{} - \braket{V_\perp(t)V_z(t')} e^{-i(\Omega_{01} t +\pi/2)}\tilde{\rho}_{01}(t) - \braket{V_\perp(t)V_z(t')}e^{+i(\Omega_{01} t+\pi/2)}\tilde{\rho}_{10}(t) \bigg\}, \\
       \dot{\tilde{\rho}}_{01}(t) = &{}+\frac{1}{\hbar^2} \int_0^{t} dt'\,\bigg\{
       2\braket{V_z(t)V_\perp(t')}e^{i(\Omega_{01} t'+\pi/2)}\Big(\tilde{\rho}_{11}(t)-\tilde{\rho}_{00}(t) \Big) \\  &{}- 2\braket{V_\perp(t)V_\perp(t')}e^{i\Omega_{01}(t-t')}\tilde{\rho}_{01}(t) + 2\braket{V_\perp(t)V_\perp(t')}e^{i(\Omega_{01} (t+t')+\pi)}\tilde{\rho}_{10}(t)  -4\braket{V_z(t)V_z(t')}\tilde{\rho}_{01}(t)\bigg\},
    \end{split}
\end{equation}
with $\dot{\tilde{\rho}}_{01}(t) = \dot{\tilde{\rho}}_{10}^*(t)$. It can be seen that, in Eqs.$\,$(\ref{liuvilleExpanded}), terms containing correlations between transverse and longitudinal components are present. However, for long times $t\gg0$ we can apply the secular approximation and neglect the fast oscillating terms, so that we obtain three uncoupled differential equations. With regard to the diagonal elements, the expression for the polarisation $\braket{\sigma_z(t)}= \frac{1}{2}\left(\rho_{11}(t)-\rho_{00}(t) \right)$ can be obtained as
\begin{equation}
\label{polarisation}
        \braket{\dot{\tilde{\sigma}}_z(t)} =  - \frac{4}{\hbar^2} \int_0^{t} \braket{V_\perp(t)V_\perp(t')}\cos\left[\Omega_{01}(t-t')\right]dt'\, \braket{\tilde{\sigma}_z(t)},
\end{equation}
while, for the coherence term $\rho_{01}(t)$ we have
\begin{equation}
\label{coherenceTerm}
    \dot{\tilde{\rho}}_{01}(t)  =   -\frac{1}{\hbar^2} \int_0^{t} dt'\,\bigg\{ 2\braket{V_\perp(t)V_\perp(t')}e^{+i\Omega_{01}(t-t')}+ 4\braket{V_z(t)V_z(t')} \bigg\}\tilde{\rho}_{01}(t).
\end{equation}
\end{widetext}
When these expressions are formally integrated, they lead to the form of the density matrix presented in Sec. \ref{sec:Noise} and to Eqs. (\ref{Gamma1genericV}) and (\ref{XiDephgenericV}). To integrate these equations, we expand $V_\perp$ and $V_z$ at first order in the noise, $V_\perp(t)\simeq \left(\partial V_\perp[x_\perp(t)]/\partial x_\perp\right)\delta x_\perp$ and $V_z(t)\simeq \left(\partial V_z[ x_z(t)]/\partial x_z\right)\delta x_z$. We also make use of the following relations between the autocorrelation $C_x(\tau)= \braket{\delta x(0)\delta x(\tau)}$ and the PSD $S_ x(\omega)$ of the stochastic variable $\delta x$:
\begin{equation}
    S_x(\omega) = \frac{1}{2\pi} \int_{-\infty}^{+\infty} C_x(\tau)e^{i\omega \tau}d\tau,
\end{equation}
 \begin{equation}
     C_x(\tau) = \int_{-\infty}^{+\infty} S_x(\omega)e^{-i\omega \tau}d\omega.
 \end{equation}
Integrating Eqs. (\ref{polarisation}) and (\ref{coherenceTerm}) yields integrals of the form $\int_0^{t} dt'' \int_0^{t''} dt'\,F(t'', t')$, with $F$ function of $t'$ and $t''$. When the integrand $F(t'', t')$ is symmetric by exchange of $t',\,t''$, we have
$$\int_0^{t} dt'' \int_0^{t''} dt'\,F(t'', t')= \frac{1}{2} \int_0^{t} dt'' \int_0^{t} dt'\,F(t'', t').$$
\newpage
This is true for $\braket{\dot{\tilde{\sigma}}_z(t)}$, and its integration leads to the form $\braket{\tilde{\sigma}_z(t)}= e^{G(t)}\braket{\tilde{\sigma}_z(0)}$ and to the Bloch-Redfield exponential decaying function. Since we assume the noise stationary, we can write $\braket{\delta x_\perp(t'')\delta x_\perp(t')}= \braket{\delta x_\perp(0)\delta x_\perp(t'-t'')}=C_{ x_\perp}(t'-t'')$ and make the change of variables 
$\tau=t'-t''$ and $T/2=t'+t''$. Using the definition of the power spectral density given above, in the limit of $t\to\infty$ we get
\begin{align}
        &{} G(t) =-\frac{2}{\hbar^2} \left|\frac{\partial V_\perp}{\partial x_\perp}\right|^2 \int_0^{t} dt'' \int_0^{t} dt'\braket{\delta x_\perp(t'')\delta x_\perp(t')} \nonumber\\ 
            &{}\hspace{30pt}\times\cos\left[\Omega_{01}(t''-t')\right] \nonumber\\
        &{}\simeq-\frac{1}{\hbar^2} \left|\frac{\partial V_\perp}{\partial x_\perp}\right|^2 t \left[ 2\pi S_{ x_\perp}(E_{01}/\hbar) + 2\pi S_{ x_\perp}(-E_{01}/\hbar) \right]\nonumber\\ 
        &{}=-\frac{4\pi}{\hbar^2} \left|\frac{\partial V_\perp}{\partial x_\perp}\right|^2 S_{ x_\perp}(E_{01}/\hbar)\,t = -\Gamma_1 t\,,
\end{align}
which corresponds to Eq.$\,$(\ref{gamma1}) in Sec. \ref{sec:Noise}. The evolution of the populations in the limit of zero temperature is thus given by $\tilde{\rho}_{11,00} = \frac{1}{2} \pm \braket{\tilde{\sigma}_z(t)} = \frac{1}{2} \pm \braket{\tilde{\sigma}_z(0)}e^{-\Gamma_1 t}$.\\
Regarding the evolution of $\tilde{\rho}_{01}$, the integration of Eq.$\,$(\ref{coherenceTerm}) yields a decaying function that is still of the form $\tilde{\rho}_{01}(t) = e^{A(t)}\tilde{\rho}_{01}(0)$, with the exponent $A(t)$ given by
\begin{align}
   &{}A(t)=  -\frac{2}{\hbar^2} \int_0^{t}dt''\int_0^{t''} dt'\, \braket{V_\perp(t'')V_\perp(t')}e^{+i\Omega_{01}(t''-t')} \nonumber\\
   &{}\qquad -\frac{4}{\hbar^2} \int_0^{t}dt''\int_0^{t''} dt'\,\braket{V_z(t'')V_z(t')}.
\end{align}
The first term in the integral is not symmetric by exchange of $t'$ and $t''$, but is can be expressed as a sum of a symmetric, real term $\propto\cos[\Omega_{01}(t''-t')]$ and an antisymmetric, imaginary one $\propto i\sin[\Omega_{01}(t''-t')]$. The latter introduces a phase factor in $\rho_{01}(t)$ and does not contribute to the dephasing. The real part can be solved in the same way as the relaxation factor was calculated above, leading to
\begin{align}
    -&{}\frac{1}{2\hbar^2} \left|\frac{\partial V_\perp}{\partial x_\perp}\right|^2  \int_0^{t}dt''\int_0^{t} dt'\, \braket{\delta x_\perp(0)\delta x_\perp(t'-t'')} \nonumber\\
    &{}\times\Big(e^{i\Omega_{01}(t''-t')} + e^{-i\Omega_{01}(t''-t')}\Big) = -\frac{\Gamma_1}{2}t.   
\end{align}
The $V_z(t'')V_z(t')$ term contributes to the pure dephasing and can be integrated as follows \cite{Ithier2005}:
\begin{align}
       -&{}\frac{2}{\hbar^2} \left|\frac{\partial V_z}{\partial x_z}\right|^2 \int_0^{t}dt''\int_0^{t} dt'\,\braket{\delta x_z(t'')\delta x_z(t')} \nonumber\\
       &{}= -\frac{2}{\hbar^2} \left|\frac{\partial V_z}{\partial x_z}\right|^2 \int_{-\infty}^{+\infty} d\omega S_{x_z}(\omega)\int_0^{t}dt''e^{i\omega t''}\int_0^{t} dt'e^{-i\omega t'} \nonumber\\
       &{}= -\frac{4}{\hbar^2} \left|\frac{\partial V_z}{\partial x_z}\right|^2  \bigg(\int_{0}^{+\infty} S_{x_z}(\omega)\,\mathrm{sinc}^2(\omega t/2)d\omega \bigg) t^2,
\end{align}
where the integral of the PSD depends on the spectral characteristics of the noise $\delta x_z$. Usually, the $1/f$ noise in superconducting systems is treated as static noise, which is valid for $\omega_{\text{IR}} \ll \omega_{\text{UV}} \ll 1/t$, with $\omega_{\text{IR}}$ and $\omega_{\text{UV}}$ the PSD cutoffs. When, in the Majorana transmon, the dephasing due to $1/f$ noise affecting the offset charge $n_g$ is numerically simulated, with the noise frequencies within the range $10\ Hz < f < 50\ MHz$, we obtain $T_2^* \sim 10 \ \mu s$, corresponding to $1/T_2^* \sim 100\ kHz$. The noise cannot be considered static in this case and another approach is needed \cite{Ithier2005}. The contribution of the function $\mathrm{sinc}^2(\omega t/2)$ is mostly restricted to $\omega< 2\pi/t$. Since $S(\omega)$ is also peaked at $\omega\sim0$, it is reasonable to restrict the integral to the frequency range $\omega_{\text{IR}}<\omega<2\pi/t$. When approximating $\mathrm{sinc}(\omega t/2)\simeq \cos^2(\omega t/2)$, we obtain
\begin{align}
    &{}\int_{0}^{+\infty} \frac{1}{\omega}\,\mathrm{sinc}^2(\omega t/2) d\omega \simeq \int_{\omega_{\text{IR}}}^{2\pi/t} \frac{1}{\omega}\,\cos^2(\omega t/2) d\omega \nonumber\\
    &{}\quad=\int_{\omega_{\text{IR}}t/2}^{\pi} \frac{\cos^2(y)}{y} dy = -\frac{1}{2}\mathrm{Ci}(\omega_{\text{IR}}t) + \frac{1}{2}\mathrm{Ci}(2\pi) \nonumber\\ &{}\quad\quad+\frac{1}{2}\mathrm{ln}(2\pi/\omega_{\text{IR}}t) \simeq  \mathrm{ln}\left(\frac{2\pi}{\omega_{\text{IR}}t}\right) + O(1),
\end{align}
where $\mathrm{Ci}(x)$ is the cosine integral. The expression for the dephasing factor is therefore given by
\begin{align}
    \left|e^{A(t)}\right|=&{}\,\mathrm{exp}\bigg\{ -\frac{2\pi}{\hbar^2}\left|\frac{\partial V_\perp}{\partial x_\perp}\right|^2 S_{ x_\perp}(E_{01}/\hbar)\,t \nonumber\\
    &{}- \frac{4}{\hbar^2} \left|\frac{\partial V_z}{\partial x_z}\right|^2 \alpha\,\mathrm{ln}\left(\frac{2\pi}{\omega_{\text{IR}}t}\right) t^2\bigg\},
\end{align}
which is the expression used for calculating the pure dephasing factor $|f_z|= e^{-\chi(t)}$ in Eq. (\ref{DephFactor}).

\input{main.bbl}  

\end{document}

%% file: main.bbl
%